\documentstyle[11pt,psfig,amssymb] {article}
\input epsf.sty
\newcommand{\be}{\begin{equation}}
\newcommand{\ee}{\end{equation}}
\newcommand{\bd}{\begin{displaymath}}
\newcommand{\ed}{\end{displaymath}}
\newcommand{\ba}{\begin{array}}
\newcommand{\ea}{\end{array}}
\newcommand{\bt}{\begin{tabular}}
\newcommand{\et}{\end{tabular}}
\newcommand{\bc}{\begin{center}}
\newcommand{\ec}{\end{center}}
\newcommand{\bn}{\begin{enumerate}}
\newcommand{\en}{\end{enumerate}}
\newcommand{\bi}{\begin{itemize}}
\newcommand{\ei}{\end{itemize}}
\newcommand{\bqr}{\begin{eqnarray}}
\newcommand{\eqr}{\end{eqnarray}}
\newcommand{\bfig}{\begin{figure}[tbp]}
\newcommand{\efig}{\end{figure}}
\newcommand{\btab}{\begin{table}[ht]}
\newcommand{\etab}{\end{tabular}\ec\end{table}}
\newcommand{\bl}{\begin{large}}
\newcommand{\el}{\end{large}}

\newcommand{\nb}{\nonumber}

\newcommand{\Ps}{P_{\sigma}}

\newcommand{\nuc}[2]{\mbox{\relax\ifmmode{}^{#1}{\protect\text{#2}}\else${}^{#1}$#2\fi}}

\title{Bare vs effective pairing forces. A microscopic finite-range \\
interaction for HFB calculations in coordinate space.
}
\author     {
\\
             T. Duguet\footnote{E-mail~: duguet@theory.phy.anl.gov} \\
\\
             {\em Argonne National Laboratory, Physics Division,} \\
             {\em 9700 South Cass Avenue, Argonne, IL 60439, USA} \\
\\
            }% end author

\begin{document}

\maketitle

\begin{abstract}

We propose a microscopic effective interaction to treat pairing correlations in the $^{1}S_0$ channel. It is introduced by recasting the gap equation written in terms of the bare force into a fully equivalent pairing problem. Within this approach, the proposed interaction reproduces the pairing properties provided by the realistic $AV18$ force very accurately. Written in the canonical basis of the actual Bogolyubov transformation, the force takes the form of an off-shell in-medium two-body matrix in the superfluid phase multiplied by a BCS occupation number $2 \, \rho_{m}$. This interaction is finite ranged, non local, total-momentum dependent and density dependent. The factor $2 \, \rho_{m}$ emerging from the recast of the gap equation provides a natural cut-off and makes zero-range approximations of the effective vertex meaningful. Performing such an approximation, the roles of the range and of the density dependence of the interaction can be disentangled. The isoscalar and isovector density-dependences derived ab-initio provide the pairing force with a strong predictive power when extrapolated toward the drip-lines. Although finite ranged and non local, the proposed interaction makes HFB calculations of finite nuclei in coordinate space tractable. Through the two-basis method, its computational cost is of the same order as for a zero-range force.



\end{abstract}

\section{Introduction}
\label{intro}

The structure of the nucleus and the properties of extended nuclear systems strongly depend on their possible superfluid nature. In finite nuclei, nucleonic pairing has a strong influence on all low-energy properties of the system. This encompasses masses, separation energies, deformation, individual excitation spectra and collective excitation modes such as rotation or vibration. The role of pairing correlations is particularly emphasized when going toward the drip-lines. This is due to the proximity of the Fermi surface to the single-particle continuum. Indeed, the scattering of virtual pairs into the continuum gives rise to a variety of new phenomena as regards the properties of ground and excited states of nuclei~\cite{doba3}.

In neutron stars, a good description of pairing is also required. The neutron-neutron $^{1}S_0$ pairing drives the physics of neutron star crusts, while at higher densities, that is from the inner crust to two or three times the saturation density, neutron and proton pairing occur dominantly in the $^{3}P_2- \, ^{3}F_2$ and $^{1}S_0$ channels, respectively~\cite{dean}. Such superfluid phases influence the dynamical and thermal evolution of the star. Indeed, post-glitching timing observations~\cite{sauls} and the cooling history~\cite{heiselberg3} strongly depend on the presence or absence of pairing in the system.

To treat pairing, one needs to specify the many-body technique used and the appropriate interaction to insert into the calculation at that chosen level of approximation. The latter depends on the situation and on the system. In the present case, our aim is eventually to perform non-relativistic self-consistent mean-field and beyond-mean-field calculations in finite nuclei. Mean-field calculations are of the Hartree-Fock-Bogolyubov (HFB) type, while the considered beyond-mean-field calculations deal typically with symmetry restorations (Projected Mean Field Method) and with large amplitude motion in nuclei (Generator Coordinate Method)~\cite{ring1,michael}. Thus, one has to identify the appropriate vertices to be used in each of these cases. The same question arises for instance in the context of the shell model~\cite{hjorthjensen}.

We concentrate here on the mean-field treatment. While the variational derivation of the HFB equations cannot help in defining the appropriate vertex, the Green function or Goldstone formalisms are able to do so. Such many-body theories show unambiguously that the irreducible vertex to be used in the pairing channel at lowest order is the bare nucleon-nucleon (NN) force~\cite{gorkov,bogo,mehta1,henley}. At the next order, the irreducible pairing vertex involves the so-called polarization diagrams. This situation is in contrast to the particle-hole channel where one cannot avoid regularizing the repulsive core of the bare interaction from the outset through the definition of an in-medium two-body matrix. This stresses the fact that the effective forces in the particle-hole and particle-particle channels are different. In particular, the direct use in the gap equation of an in-medium vertex such as the Brueckner $G$-matrix leads to double counting~\cite{henley,baldo3}. 

Thus, the mean-field energy defined in this context is a functional of one-body normal and abnormal density matrices and do not refer to the mean value of a given Hamiltonian in a product state. However, the strategy used consists of motivating the low-energy functional from a many-body expansion and thus, consists of keeping an explicit link to the bare $NN$ force. With this approach, the theory is not necessarily local and the mean-field functional does not include more correlations than provided by the irreducible vertices in the particle-hole and particle-particle channels. Also, one has to come back to the many-body expansion to enrich the functional if going beyond the mean-field approach~\cite{duguet9}. Even if the final goal is the same, this differs from the strategy used in the so-called (local) density functional theory~\cite{kohn1,kohn2}. Note also that calculations based on the extension of Gorkov's formalism to the relativistic case are also performed~\cite{kucharek1,brett1,brett2}. Recently, the effects of polarization leading to the screening of nucleon and meson propagators have been studied in infinite matter~\cite{chen}. The reduction of the gap was found to be much larger than in the non-relativistic case, correcting to some extent for the excessive gap found at first order.

As regards HFB calculations in finite nuclei, only phenomenological forces have been used in the spin-singlet/isospin-triplet pairing channel so far. One example is the finite-range, density-independent Gogny force~\cite{dech}, whose restriction to the spin-singlet/isospin triplet channel can be written as:

\begin{equation}
V^{Gogny}_{\tau} (\vec{r}_1, \vec{r}_2) \, =  \, \frac{1 - P_{\sigma}}{2} \, \, \sum_{i=1}^{2} \, \, \lambda^{i}_{\tau} \,  \,  e^{-|\vec{r}_1 - \vec{r}_2|^2/\alpha^2_i} \, \, \, \, \, ,
\label{gogny}
\end{equation}
and is to be averaged over the angle between the incoming and outgoing relative momenta if dealing with the $S$ wave only. In Eq.~\ref{gogny}, $P_{\sigma}$ is the spin-exchange operator while the Coulomb part of the Gogny force has not been considered. The other commonly used pairing interaction is the (density-dependent) delta interaction (DDDI)~\cite{bertsch,rigol,duguet2,duguet6,duguet61}: 

\begin{equation}
V^{^{1}S_0}_{\tau} (\vec{r}_1, \vec{r}_2) \, = \, \lambda_{\tau} \, \,  \frac{1 - P_{\sigma}}{2} \, \, \left[1 - \left(\frac{\rho(\frac{\vec{r}_1 + \vec{r}_2}{2})}{\rho_{c}}\right)^{\gamma} \, \right] \, \, \delta (\vec{r}_1 - \vec{r}_2 ) \, \, \, \, \, .
\label{DDDI}
\end{equation}
where $\rho$ denotes the matter density (= local scalar-isoscalar part of the one-body density-matrix).

The latter, usually used when solving the problem in coordinate space, must be complemented with a cut-off in the gap equation to avoid divergences. Studies of the rotational bands of super-deformed nuclei~\cite{terasaki2,rigol} and actinides~\cite{duguet2,michael1}, of halo nuclei~\cite{bertsch}, and of the evolution of charge radii across magic numbers~\cite{tajima2,fayans2} have helped to establish the success and the surface-peaked character of the DDDI in the pairing channel. Recently, more systematic studies of asymptotic matter and pair densities of exotic nuclei~\cite{doba4}, of the evolution of the pairing gap toward the neutron drip-line~\cite{doba2}, and the average behavior of the odd-even mass differences over the mass table~\cite{doba7} have allowed a refinement of the DDDI. The optimal compatibility between experimental data and mean-field calculations was obtained for a force between surface and volume~\cite{duguet2,doba7}, with $\rho_c \approx 2 \rho_{sat}$, where $\rho_{sat}$ = 0.16 fm$^{-3}$ is the saturation density of symmetric nuclear matter. The great sensitivity of matter and pair densities to pairing in the low-density regime seemed to favor $1/2 \leq \gamma \leq 1$.

Although successful in describing low-energy nuclear structure over the (known) mass table~\cite{michael,rigol,duguet2,fallon,ph,goriely1,samyn,goriely2}, the two previous phenomenological pairing interactions lack a link to the bare nucleon-nucleon interaction. They were directly fitted to finite nuclei data, and may thus renormalize beyond-mean-field effects. In addition, their fits were performed where experimental data are available. Extrapolating the use of these interactions toward the drip-lines is questionable. To correct for this defect, few DDDI were fitted to reproduce the gap provided by realistic $NN$ forces in infinite matter~\cite{garrido1,garrido2}. However, the necessary density dependence and cut-off were still treated phenomenologically.

In fact, the present knowledge about the pairing force and the nature of pairing correlations in nuclei is quite poor. Properties such as the range of the effective pairing interaction, its link to the bare force, its possible surface character in finite nuclei and its density dependence, in particular isovector, still have to be clarified. As noticed as early as thirty years ago~\cite{migdal} and pointed out several times since~\cite{doba3,duguet9,fayans2,doba2,doba7,vauth,doba8,duguet8}, obtaining proper density dependences of particle-hole and particle-particle effective forces at a given level of approximation is difficult but of great importance to meet modern high-precision experiments. The problem dealing with the cut-off to be used in the pairing channel in connection with zero-range vertices has been solved recently~\cite{bulgac1,bulgac2}. The idea was to identify the divergences stemming from the use of a local gap and regularize them through a well-defined renormalization scheme. That scheme has to be understood in the context of the local density functional theory~\cite{fayans2} (which is perfectly fine) rather than as a mean-field approximation arising at lowest order of some many-body expansion. 

The present work concentrates on the mean-field treatment of pairing at relatively low energy and low density in the isotropic, spin-singlet and isospin-triplet channel. The aim of this study is manifold. In section~\ref{definition1}, we define an appropriate simple version of the bare force in the $^{1}S_0$ channel and explain in details its fitting procedure. Then, an in-medium pairing interaction (Eq.~\ref{effectv}) equivalent to the bare force is introduced in section~\ref{Formalism}. The corresponding diagrammatic resummation authorizes the study of the finite-ranged effective force and of its zero-range approximation on the same footing. This is discussed in section~\ref{Infinitematter} where the roles of the range and of the density dependence of the pairing interaction are disentangled. In the particular case of the zero-range approximation for the vertex, the scheme proposed presents strong similarities with the regularization procedure introduced in Refs.~\cite{bulgac1,bulgac2}. We will actually discuss this particular point in a forthcoming publication. Ultimately, the interaction is to be used in calculations performed in coordinate space by solving the HFB equations on a three dimensional mesh~\cite{doba6,gall,terasaki3}. Although finite ranged and non local, the proposed interaction is shown in section~\ref{interactionrealspace} to make these calculations tractable. The formulas defining completely the new effective pairing force can be found in the same section. Some important points are discussed in section~\ref{Discussion} while our conclusions are given in section~\ref{Conclusions}.

\section{Simplified bare force}
\label{definition1}

\subsection{Fitting procedure}
\label{Adjustment}

Screening effects beyond the mean-field approximation due to density and spin fluctuations are known to strongly decrease the pairing gap in neutron matter, both for singlet and triplet pairing~\cite{dean,clark2,baldo,shen,schulze2}. Whether it is justified to extend this statement to finite nuclei is still an open question. Indeed, the dressing of the vertex could change significantly when going from neutron matter to symmetric matter~\cite{heiselberg4}. Also, including the induced interaction and off-shell self-energy effects due to the exchange of surface vibrations between time-reversed states seems to increase the pairing gap in finite nuclei compared to that generated by the bare force~\cite{terasaki2,terasaki1,barranco}. In addition, the influence of the restoration of particle-number and pairing vibrations still have to be characterized from systematic self-consistent calculations in even and odd nuclei. The situation as regards beyond-mean-field effects is unclear at this stage and one can simply state that a significant cancellation between the effects of screening and of surface vibrations on singlet pairing should take place in finite nuclei. As a result, two strategies seem reasonable when dealing with the definition of a pairing interaction to be used in mean-field calculations.

Sticking to the pure mean-field picture, one can define the interaction by reproducing properties of the bare $NN$ interaction. If necessary, the possibility remains to include beyond-mean-field effects explicitly in a consistent way when using an interaction mimicking the bare force.

A second strategy consists of fitting the interaction directly to finite nuclei data through mean-field calculations~\cite{rigol}. This strategy has been the most popular so far when dealing with phenomenological forces to be used in self-consistent Hartree-Fock-Bardeen-Cooper-Schrieffer (HFBCS) and HFB calculations~\cite{dech,rigol}. Such a procedure aims at renormalizing the beyond-mean-field effects which possibly do not cancel out. Of course, the use of such a force in calculations going explicitly beyond the mean-field is suspicious.

As a first attempt, and because we want to separate mean-field from beyond-mean-field effects, we follow the first strategy. Note that no isospin symmetry breaking effects due to Coulomb or charge-independence breaking of the nuclear part of the interaction is considered in the present study. It was shown to have no effect on the gap in the $^{1}S_{0}$ channel~\cite{elgar1}. Also, we do not consider the neutron-proton component of the force in this channel.

\subsection{Form of $V^{^{1}S_{0}}_{sep}$}
\label{subdefinition1}

As already said, the bare $NN$ interaction has to be considered in the pairing channel at lowest order in irreducible vertices. However, the full complexity of any realistic $NN$ force makes systematic HFB calculations in finite nuclei untractable from the computational point of view. We thus have to define a simplified bare force retaining the essential physics provided by the full $NN$ interaction as regards pairing.

A particular feature of the $NN$ force in the $^{1}S_0$ channel is the corresponding very large, negative scattering length. The empirical values for neutron-neutron, neutron-proton and proton-proton scattering length are $a^{^{1}S_0}_{nn} = - 18.5 \pm 0.4 \, \,  fm$~\cite{teramond} ($- 18.7 \, \pm 0.6 \, \, fm$ in a recent experiment~\cite{gonzales}), $a^{^{1}S_0}_{np} = - 23.749 \pm 0.008 \, \,  fm$~\cite{koester} and $a^{^{1}S_0}_{pp} = - 7.8063 \pm 0.0026 \, \,  fm$~\cite{bergervoet} respectively. This indicates that the $NN$ interaction holds a virtual state in the vacuum at almost zero scattering energy in the $^{1}S_0$ channel. In the vicinity of the virtual state, the scattering $t$-matrix can be written in a separable form, suggesting that the bare interaction itself is to a good approximation separable and non-local at low energy~\cite{brown}. Thus, we start from the definition of the interaction $V^{^{1}S_{0}}_{sep}$ in the plane-wave basis 

\begin{equation}\label{eq1}
\phi_{\vec{k}_is_i}(\vec{r}) = \; \langle \, \vec{r} \; | \; \vec{k}_is_i \, \rangle  \; =
e^{i\vec{k}_i.\vec{r}} \;
\chi^{s_i}_{\frac{1}{2}} \, \, \, ,
\end{equation}
through

\begin{eqnarray}
\langle \, \vec{k}_{1} s_{1} \vec{k}_{2} s_{2} \, |  \, \frac{1-\Ps}{2} \, V^{^{1}S_{0}}_{sep} \, |\, \vec{k}_{3} s_{3} \vec{k}_{4} s_{4} \rangle_{a} \!  &=&  \! \frac{1}{2} \, \langle \, \vec{k}_{1} \vec{k}_{2} \, |  \, V^{^{1}S_{0}}_{sep} \, |\, \vec{k}_{3} \vec{k}_{4} \rangle_{s} \, \left(\delta_{s_{1}s_{3}} \, \delta_{s_{2}s_{4}} \! -  \! \delta_{s_{1}s_{4}} \,\delta_{s_{2}s_{3}}\right) \nb \\
&& \label{matrixelement} \\
\langle \, \vec{k}_{1} \vec{k}_{2} \, |  \, V^{^{1}S_{0}}_{sep} \, |\, \vec{k}_{3} \vec{k}_{4} \rangle \!  &=&  \! \langle \, \vec{k} \, |  \,  V^{^{1}S_{0}}_{sep} \, | \, \vec{k} \,' \, \rangle \, (2\pi)^{3} \, \delta (\vec{P}-\vec{P} \,') \, \, \, \, , \nb
\end{eqnarray}
where its center of mass part is approximated by:

\begin{equation}
\langle \, \vec{k} \, |  \,  V^{^{1}S_{0}}_{sep} \, | \, \vec{k} \,' \, \rangle \, = \, \lambda \, v(k) \, v(k') \, \, \, .
\label{approxmatrixelement}
\end{equation}

While $\vec{k}_{i}$ denotes the momentum of a particle in the laboratory frame, $\vec{P}=\vec{k}_{i}+\vec{k}_{j}$ and $\vec{k}=(\vec{k}_{i}-\vec{k}_{j})/2$ are the total and relative momenta of a pair, respectively. The states $\{| \; \vec{k}_i s_i \, \rangle \}$ span the tensor product of momentum and spin ($s_i = \pm \frac{1}{2}$) single-particle Hilbert spaces and are orthonormalized through\footnote{Because of the convention used to define plane waves, integrals in momentum space are characterized by $\int d^{3} \vec{k} \, / \, (2\pi)^{3}$. We also use the convention $\hbar^{2} = 1$.}

\begin{equation}
\langle \, \phi_{\vec{k_i}s_i} \: | \: \phi_{\vec{k_j}s_j} \, \rangle \; = 
\; (2\pi)^3 \, \delta(\vec{k_i}-\vec{k_j}) \; \delta_{s_is_j} \, \, .
\label{orthonormalization}
\end{equation}

Also, the subscripts $a$ and $s$ in Eq.~\ref{matrixelement} denote antisymmetrized and symmetrized matrix elements respectively. A matrix element with no subscript is neither symmetrized nor antisymmetrized. The isospin quantum number has not been specified since the form of the matrix elements in the $\{T=1, T_{z}=\pm1\}$ channels is trivial.

\subsection{Connection with scattering phase-shifts}
\label{subdefinition2}

The link between the bare force and the $NN$ phase-shifts is obtained by treating the two-body problem in the center of mass frame. To make this link, it is convenient to introduce the energy-dependent scattering $t$-matrix. The Lippmann-Schwinger equation~\cite{brown} defining it in the uncoupled $^{1}S_{0}$ channel takes the form:

\begin{equation}
\langle \, \vec{k} \, |  \,  t^{^{1}S_{0}} (s) \, | \, \vec{k} \,' \, \rangle \, = \, \langle \, \vec{k} \, |  \,  V^{^{1}S_{0}} \, | \, \vec{k} \,' \, \rangle \, + \, \int \, \frac{d^{3} \vec{k} \,''}{(2\pi)^{3}} \, \langle \, \vec{k} \, |  \,  V^{^{1}S_{0}} \, | \, \vec{k} \,'' \, \rangle \, F^{\, t}_{\vec{P} \, \vec{k} \,''} (s) \, \langle \, \vec{k} \,'' \, |  \,  t^{^{1}S_{0}} (s) \, | \, \vec{k} \,' \, \rangle \, \, \, .
\label{offshellt}
\end{equation}
where $\vec{P}$ is the conserved total momentum of the pair and

\begin{equation}
F^{\, t}_{\vec{P} \, \vec{k}} (s) \, = \, \frac{1}{s - {k}^2 / m + i \epsilon} \, \, \, ,
\label{2bodypropvac}
\end{equation}
is the non-interacting two-particles propagator in free-space appropriate for outgoing boundary conditions. In Eq.~\ref{2bodypropvac}, $m$ is the nucleon bare mass and $s$ is the energy of the interacting pair in its center of mass (total energy subtracted by $P^{2}/4m$). It is worth noting that $F^{\, t}$ is independent of $\vec{P}$ and diagonal in $\vec{k}$.

Because of the rank-one separable form chosen for the bare force, the Lippmann-Schwinger equation is exactly solvable and $t^{^{1}S_{0}}$ takes for any triplet ($s,k,k'$) the separable form~\cite{brown}:

\begin{equation}
\langle \, \vec{k} \, |  \,  t^{^{1}S_{0}} (s) \, | \, \vec{k} \, '  \, \rangle \, = \, \lambda \, v(k) \, v(k') \, / \, \left[1 - \lambda \, \int_{0}^{\infty} \frac{d \, k''}{2\pi^2} \, \frac{k'' \, ^2 \, v^2(k'')}{s-k'' \, ^2/m + i \epsilon} \right] \, \, \, .
\label{solutiontmatrix}
\end{equation}

While the full scattering problem  is expressible in terms of the half on-shell $t$-matrix, the phase shifts carrying the information about the two-body wave-function at long distances relate to the fully on-shell part of $t$ ($s = {k}^2/m = {k'}^2/m$). With the convention chosen to define the plane-wave basis, this link can be written explicitely under the form:

\begin{equation}
\langle \, \vec{k} \, |  \,  t \, ({k}^2/m) \, | \, \vec{k} \, '  \, \rangle \, = \,  - \frac{4 \pi}{ m k} \, \sum_l \, (2 l + 1) \, e^{i  \, \delta_l \, (k)}\sin \delta_l (k) \, P_l \, (\cos \hat{k}.\hat{k} \, ')  \, \, \, ,
\label{phaseshift1}
\end{equation}
where $\delta_l (k)$ denotes the phase shifts for a relative orbital angular-momentum $l$. We have not considered in Eq.~\ref{phaseshift1} the coupling between different $l$ channels as provided by the tensor force. Focusing on $l=0$ and using the separable $t$-matrix, one gets:

\begin{equation}
\tan  \delta^{^{1}S_{0}} \, (k) \, = \,   - \lambda \, m \, \pi \, k \, v^2(k) \, /  \, \left[\pi^2 - 2 \lambda m \, \, {\cal P} \int_{0}^{\infty} d \, k'' \, \frac{k'' \, ^2 \, v^2(k'')}{k^2-k'' \, ^2} \right]   \, \, \, , 
\label{phaseshift2}
\end{equation}
where ${\cal P}$ denotes the principal value of the integral.

Since the separable form of the bare interaction in the $^{1}S_0$ channel is physically motivated, one can hope to define an efficient and simple enough interaction by plugging the known phase-shifts into Eq.~\ref{phaseshift2} to fix its parameters. Indeed, it has been shown that the $^{1}S_0$ gap in nuclear matter is entirely determined by the $NN$ scattering phase-shifts in the vacuum~\cite{elgar1,emery2}. Of course, a rank-one separable interaction cannot reproduce the phase shifts up to infinite energy in this channel since they change sign around $250 \, MeV$~\cite{elgar1,kwong}. One could use a rank-two separable form to take care of this~\cite{kwong}. However, an overall reproduction of the phase shifts up to $E_{lab} = 250 \, MeV \Leftrightarrow k = 1.73 \, fm^{-1}$ should be sufficient to describe pairing at relatively low density. Indeed, the kernel of the gap equation is strongly peaked at $k_{F}$.

The inverse scattering problem, which corresponds to the determination of a two-particle potential from the knowledge of the phase shifts at all energies, is exactly and uniquely solvable for rank-one separable potentials~\cite{brown}. Thus, given the phase shifts, a unique solution exists for $v (k)$. We do not proceed however through the resolution of the inverse scattering problem. Indeed, a simple analytic form of $v(k)$ will eventually be necessary to perform HFB calculations of finite nuclei. We choose a simple form for $v (k)$ and try to reproduce $\delta^{^{1}S_{0}} \, (k)$ as well as possible. We consider the Gaussian form:

\begin{equation}
v (k) \, = \, e^{-\alpha^{2} k^{2}}   \, \, \, ,
\label{formofv}
\end{equation}
where $\alpha$ is the second parameter of the force. This choice will be shown later to be appropriate.

\subsection{$^{1}S_{0}$ pairing gap}
\label{pairinggap}

We have introduced a simple force to mimic the realistic $NN$ interaction in the $^{1}S_{0}$ channel. Starting from this interaction defined in the vacuum, one can go to the medium and compute the pairing gap through the BCS gap equation~\cite{bardeen}\footnote{Solving the gap equation using a separable force was done as early as 1964~\cite{kennedy}. In that work, qualitatively similar results as those we derive in infinite matter were obtained using Yukawa type interactions.}. This scheme corresponds to the lowest order in the Goldstone-Brueckner-Bogolyubov perturbation theory~\cite{mehta1,henley}, or in the Galitskii-Gorkov Green function method~\cite{gorkov,galitskii}, and defines a meaningful mean-field picture. The simplest medium to consider at this stage is infinite nuclear matter. Indeed, its translational invariance strongly simplifies the treatment and avoids the additional effects associated with the finiteness of the nucleus. Of course, it is to some extent a toy model, even if neutron stars can be considered as being closely connected with it.

\begin{figure}
\begin{center}
\leavevmode
\centerline{\psfig{figure=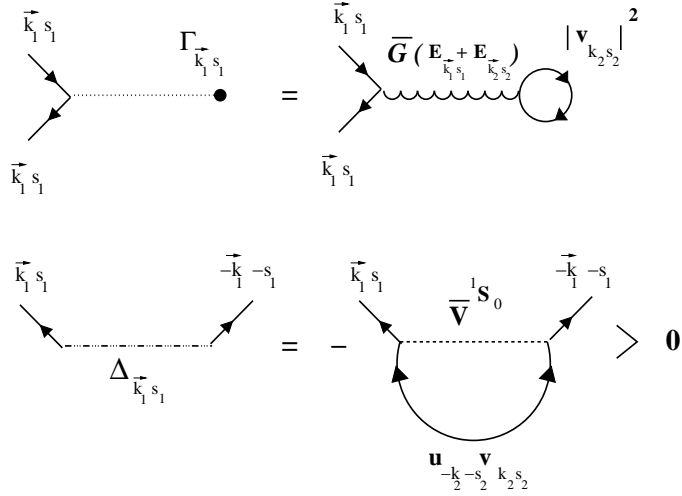,height=6.5cm}}
\end{center}
\caption{Definition of the normal $\Gamma_{\vec{k}_{1}}$ and abnormal $\Delta_{k_{1}}$ parts of the self-energy at lowest order in irreducible vertices. Combining these definitions with the usual expressions of the BCS occupation numbers $u^{2}_{k_{i}}$ and $v^{2}_{k_{i}}$, obtained from the compensation of the so-called {\it dangerous diagrams} at lowest order, provides the BCS gap equation~\cite{bogo,mehta1,henley}. The irreducible kernel entering the definition of $\Gamma_{\vec{k}_{1}}$ at lowest order is the $\, {\cal G}$-matrix summing particle-particle ladders. The two-body propagator characterizing the intermediate states in the ladder is the product of one-body mean-field Green functions defined in the superfluid system. The irreducible kernel entering the definition of $\Delta_{k_{1}}$ at lowest order is the bare $NN$ force. The reason why the $\, {\cal G}$-matrix or the $\, {\cal T}$-matrix cannot enter the gap equation originates from the necessity to avoid double-counting when compensating for the dangerous diagrams. It requires the exclusion of isolated particle-particle and hole-hole intermediate states in those vacuum-to-pair diagrams involving an abnormal contraction.}
\label{diagrams}
\end{figure}

In infinite matter, the favored Bogolyubov transformation correlates pairs of nucleons with zero total-momenta\footnote{Some exceptions exist however. See Refs.~\cite{larkin,fulde,sedrakian}.}. The gap equation written in the plane-wave basis is of the usual BCS form and reads in the $^{1}S_{0}$ channel as:

\begin{equation}
\Delta_{\vec{k},-\vec{k}} \, \equiv \, \Delta_{k} \, = \,  - \int_{0}^{\infty}  \frac{d \, k'}{ 2\pi^{2}}  \, k' \, ^2 \,  \langle \, \vec{k} \, |  \,  V^{^{1}S_{0}}_{sep} \, | \, \vec{k} \, ' \, \rangle \, \frac{\Delta_{k'}}{2 E_{k'}}   \, \, \, ,
\label{gapequation1}
\end{equation}
where $E_{k} = \sqrt{(\epsilon_{k} - \mu)^{2} + \Delta^{2}_{k}}\,$ is a quasi-particle energy; $\epsilon_{k}$ being the in-medium on-shell single-particle energy associated with the state $\phi_{\vec{k}}$ and $\mu$ the chemical potential. To be consistent, $\mu$ should be calculated iteratively by constraining the density of the system, while $\epsilon_{k}$ should be defined after regularizing the repulsive core of the bare force and by taking the influence of pairing correlations explicitly into account. This would typically require the use of the on-shell self-energy $\Gamma_{\vec{k}_{1}}$ computed from the Galitskii $\, {\cal T}$-matrix or the Brueckner $\, {\cal G}$-matrix defined in the superfluid phase~\cite{mehta1,henley,bozek}. A corresponding mean-field scheme is depicted diagrammatically on Fig.~\ref{diagrams}.  Such a procedure is very involved, especially if using modern realistic $NN$ forces in all $(S,T)$ channels. It becomes prohibitive when dealing with finite nuclei. Thus, approximate schemes making use of the Brueckner $G$~\cite{brueck} or the Feynman-Galitskii $T$~\cite{galitskii} matrices in the normal fluid are usually considered in infinite matter.

When performing extensive mean-field calculations of finite nuclei, a phenomenological effective vertex such as the Gogny~\cite{dech} or the Skyrme~\cite{vauth,skyrme} force is employed in the particle-hole channel to approximate one of the previous in-medium matrices. They usually incorporate additional phenomenology (and to some extent the effect of higher order terms) by fitting some experimental data at the mean-field level. As we are not looking for refined calculations at this stage, and because we want to use a clean theoretical quantity to adjust the separable interaction, we simply insert free single-particle energies $\epsilon_{k} = k^{2}/2m$ into Eq.~\ref{gapequation1}. We also take $\mu$ to be equal to $k_{F}^{2}/2m$, where $k_{F} = (3\pi^{2}\rho)^{1/3}$ is the Fermi momentum of one kind of nucleons in the free gas at the density $\rho$.

Considering these approximations and inserting the separable interaction into Eq.~\ref{gapequation1}, the solution of the gap equation takes the form $\Delta_{k} = \Delta_{0} \, v(k)$, with the gap at zero momentum $\Delta_{0}$ satisfying the equation~\cite{elgar1}:

\begin{equation}
1 \, = \,  - \int_{0}^{\infty}  \frac{d \, k}{ 4\pi^{2}}  \, \frac{\lambda \, k^2 \,  v ^2(k)}{\sqrt{(k^{2}/2m - k_{F}^{2}/2m)^{2} + \Delta^{2}_{0} \, v ^2(k)}}   \, \, \, .
\label{gapequation2}
\end{equation}

After solving the Eq.~\ref{gapequation2}, the gap at the Fermi surface is obtained through $\Delta_{k_{F}} = \Delta_{0} \, v (k_{F})$.

\subsection{Fit}
\label{results}

We perform a combined fit of the force on the neutron-neutron scattering phase-shifts and on the pairing gap in infinite matter provided by the modern $AV18$ $NN$ interaction~\cite{wiringa2}. Among other features, $AV18$ fits the proton-proton and neutron-proton phase shifts up to 350 MeV, as well as neutron-neutron low-energy parameters (scattering length and effective range) in the $^{1}S_{0}$ channel. Thus, the neutron-neutron phase-shifts that we use beyond the validity of the effective range approximation ($\Leftrightarrow k \geq 0.2 \, fm^{-1}$) are state of the art theoretical predictions.

\begin{figure}
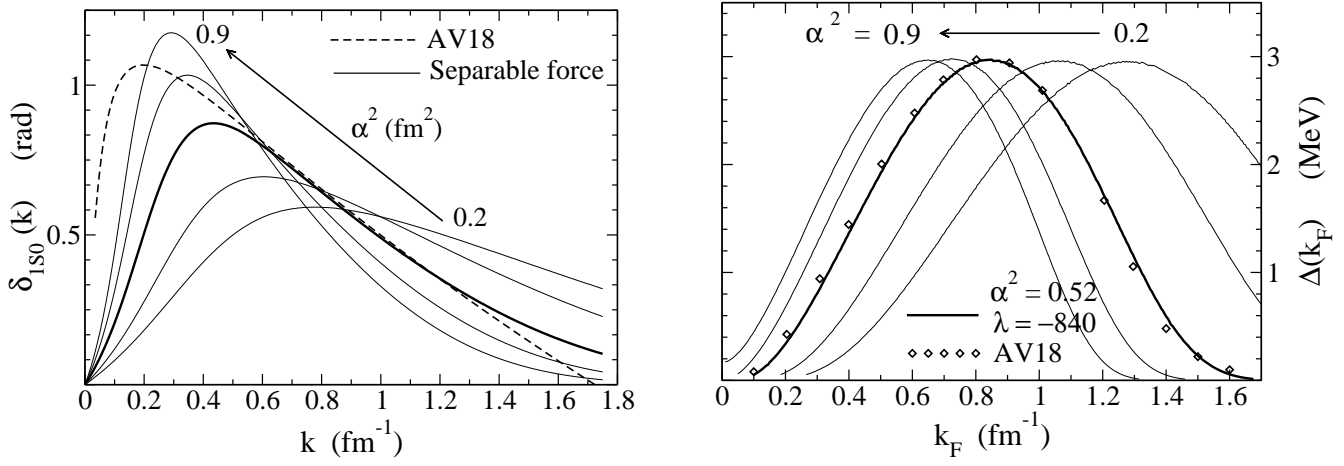

\begin{center}
\leavevmode
\centerline{\psfig{figure=phaseshiftalpha.eps,height=6cm} \hspace{0.8cm} \psfig{figure=gaps3MeV.eps,height=6.1cm}}
\end{center}
\caption{Left panel: comparison between the neutron-neutron $^{1}S_{0}$ scattering phase-shifts obtained from $AV18$ and from the rank-one separable force for several sets of parameters $(\lambda,\alpha)$. Right panel: same comparison for the $^{1}S_{0}$ pairing gap. In both cases, the range of the force is varied while the intensity is chosen accordingly to obtain a maximum gap at 3 $MeV$.}
\label{gap1}
\end{figure}

Thus, Eqs.~\ref{phaseshift2} and~\ref{gapequation2} are solved for several set of parameters $(\lambda,\alpha)$. The results are compared with those derived from $AV18$ with the same $\epsilon_{k}$. The gap $\Delta_{k_{F}}$ is plotted on the right panel of Fig.~\ref{gap1} for several values of the range $\sqrt\alpha$; the intensity $\lambda$ being chosen in order to obtain the maximum at $3 MeV$. Indeed, this is a solid prediction from all modern realistic forces~\cite{baldo3,elgar1,khodel,lombardo1}. Note that the $^{1}S_0$ gap calculated with free kinetic energies is very similar for all modern forces~\cite{baldo3,elgar1,khodel,lombardo1}. This is due to the fact that they all reproduce the phase shifts very accurately. One can see that the gap strongly depends on $\alpha$. As the pairing gap probes the interaction in a very sensitive way, requiring the precise reproduction of $\Delta_{k_{F}}$ derived from $AV18$ allows little latitude and determines the parameters of the separable force quite uniquely. This is a nice feature. For the best set of parameters ($\lambda = -840 \, MeV.fm^{3}$, $\alpha = \sqrt{0.52} \,  fm$), $\Delta_{k_{F}}$ is reproduced almost perfectly up to the gap closure. In particular, the bump is obtained at the right density and energy. This is a non trivial result in view of the very simple form of our bare force. The success of the procedure comes back to the justification of its separable form and to the overall reproduction of the phase shifts. 

The phase shifts calculated using the same sets of parameters are compared to those predicted by $AV18$ on the left panel of Fig.~\ref{gap1}. The simplicity of the force used seems to be more critical as regards a precise reproduction of the phase shifts. Also, this quantity hardly constraints the parameters in an obvious way. This justifies the complementary use of the pairing gap to fit the interaction. Interestingly enough, the gap is well reproduced at very low density when using our best set of parameters (thick curve on both panels), while the phase shifts are poorly reproduced below $k = 0.4 \, fm^{-1}$. In particular, not reproducing the large scattering length $a^{^{1}S_0}_{nn}$ does not seem to be a major problem to obtain excellent gaps, as long as $\delta^{^{1}S_0} (k)$ is correctly treated beyond $k = 0.4 \, fm^{-1}$ as seen in the left panel of Fig.~\ref{gap1}. It is known that concentrating on the very low energy part of the phase shifts (= effective range approximation) allows a good reproduction of the gap up to $k_{F} =  0.5 \, fm^{-1}$ but fails badly beyond that point~\cite{elgar1}. As long as the quasi-bound state exists in the $^{1}S_0$ channel to motivate the separable form of the force, it seems non-essential to obtain a precise value of its eigenenergy (related to  $a^{^{1}S_0}_{nn}$, that is to the slope of $\delta^{^{1}S_0} (k)$ at $k=0$). It seems more important to get an overall reproduction of $\delta^{^{1}S_0} (k)$ which, through the on-shell $t$-matrix, relates to the wave-function of the virtual state at intermediate distances~\cite{brown}. This result agrees with the conclusions of Ref.~\cite{khodel} and balances those obtained in the relativistic context~\cite{brett2}.

\subsection{Analysis of $V^{^{1}S_{0}}_{sep}$}
\label{Analysis}

\begin{figure}
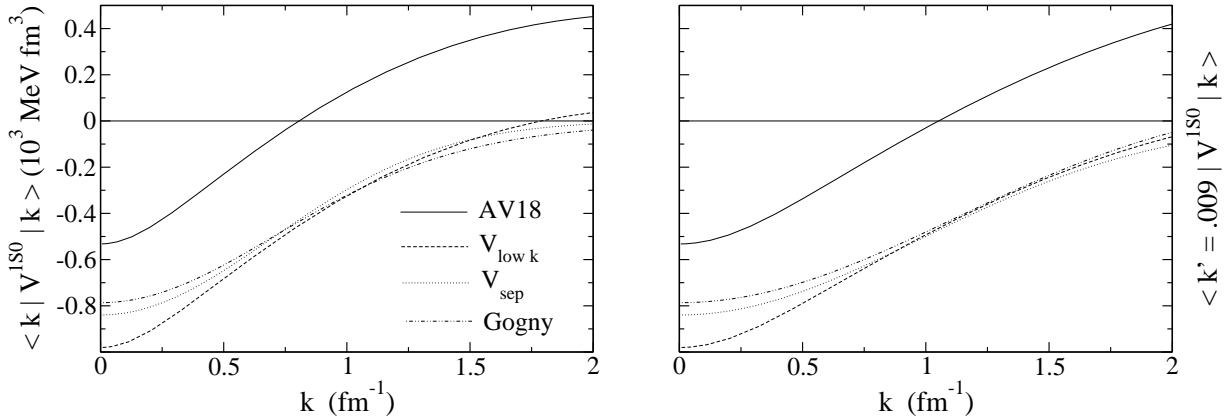

\begin{center}
\leavevmode
\centerline{\psfig{figure=v1S0diago.eps,height=5.5cm} \hspace{0.7cm} \psfig{figure=v1S0nondiago.eps,height=5.5cm}}
\end{center}
\caption{Matrix elements in momentum space of $AV18$ (full line), $V_{low \, k}$ (dashed line), our separable force (dotted line) and the Gogny force with the D1S parametrization~\cite{d1s} (dashed-dotted line) in the $^{1}S_0$ channel. Left panel: diagonal matrix elements. Right panel: non-diagonal matrix elements at $k' = 0.009 \, fm^{-1}$.}
\label{comparisonvlowk}
\end{figure}

As already mentioned, the rank-one separable form does not allow the description of the negative part of the $^{1}S_0$ phase shifts at high energy. As seen on the left panel of Fig.~\ref{gap1}, this translates into an overshoot of the phase shifts beyond $k = 1.4 \, fm^{-1}$, but only into a slight over estimation of the gap at $k_F = 1.4 \, fm^{-1}$. Roughly speaking, the justification for not resolving explicitly the hard core of the realistic bare force is similar to the one providing the grounds for the $V_{low \, k}$ interaction~\cite{bogner} or motivating the description of nuclear matter through effective field theories~\cite{steele1,steele2}. Of course, we only take care of the $^{1}S_0$ channel here, we do not integrate out the high relative momentum components of the bare force explicitly and thus, we do not look for a high precision potential model. In Fig.~\ref{comparisonvlowk}, the matrix elements of our separable interaction in momentum space are compared with those of $AV18$ and those of the non-hermitian $V_{low \, k}$ obtained from $AV18$~\cite{bogner2}. The cut-off used for $V_{low \, k}$ is $\Lambda = 2.1 \, fm^{-1}$. Our $V^{^{1}S_{0}}_{sep}$ is very close to $V_{low \, k}$ and quite different from $AV18$ itself\footnote{$V^{^{1}S_{0}}_{sep}$ and $V_{low \, k}$ differ from $AV18$ by a constant shift in momentum space equivalent to a contact term in coordinate space. This contact term properly deals with the short range part of the $NN$ interaction at low energy.}. It clarifies the physical content of our separable interaction and characterizes our scheme as a low-energy effective theory of nuclear matter valid below $\approx 3 \, \rho_{sat}$. Note that, while the matrix elements of modern realistic forces are scattered, their $V_{low \, k}$ partners all look the same~\cite{bogner2}. In the same way, the previous fitting procedure would lead to very similar separable interactions by starting from other modern realistic forces.

As seen in Fig.~\ref{comparisonvlowk}, the range and the non locality of $V^{^{1}S_{0}}_{sep}$ are of very good quality while its intensity at small $k$ characterized by $\lambda = - 840 \, MeV.fm^{3}$ is slightly too low. It originates from the poor description of the phase-shifts below $k = 0.4 \, fm^{-1}$, that is, from missing the virtual state at almost zero energy. This could not be avoided because of the simple Gaussian form used for $v (k)$. Note that keeping the range fixed, a pole at zero energy would have been obtained in the $t$-matrix derived from $V^{^{1}S_{0}}_{sep}$ for $\lambda = - 950 \, MeV.fm^{3}$. Last but not least, the $^{1}S_0$ gaps obtained in infinite matter with our force and with $V_{low \, k}$ are quite similar~\cite{sedrakian1}. However, as pairing is exponentially sensitive to the force intensity, the slightly stronger diagonal matrix elements of $V_{low \, k}$ up to $k = 1.5 \, fm^{-1}$ translate into a slight overshoot of the gap provided by AV18 or by our force~\cite{sedrakian1}.

The Gogny interaction can be considered as a benchmark concerning pairing properties in finite nuclei~\cite{michael,dech,doba8,anguiano,peru,egido1,girod}. As regards the calculation of the $^{1}S_0$ gap in infinite matter, it was shown to behave almost like a bare force, especially when using the D1S~\cite{d1s} parametrisation~\cite{garrido1,garrido2}. To address more precisely whether the Gogny force mimics the bare force in the $^{1}S_{0}$ channel, its diagonal and non-diagonal matrix elements~\cite{kucharek2} are also plotted on Fig.~\ref{comparisonvlowk} for the D1S parametrization~\cite{d1s}. The Gogny force appears to be similar to the $V_{low \, k}$ interaction and to our separable bare force. The underestimate at very low momenta is of no importance for pairing as discussed for $V^{^{1}S_0}_{sep}$. The similarity with $V_{low \, k}$ and $V^{^{1}S_0}_{sep}$ seems to explain why the Gogny force provides equivalent gaps to those obtained from the bare force. Looking more into detail, a slight overshoot of the gap provided by $V_{low \, k}$, and thus by AV18, is obtained with Gogny beyond $k_{F} = 0.8 fm^{-1}$~\cite{garrido1,sedrakian1}. This can be related to its slightly too large diagonal matrix elements compared to those of our separable force beyond $k = 0.75 \, fm^{-1}$. As densities beyond $k_{F} = 0.8 fm^{-1}$ dominate in finite nuclei, such an overshoot could be sufficient to explain why the bare force would not provide enough pairing in these systems and characterize the necessary beyond-mean-field effects. We will come back to this in section~\ref{Discussion}. Let us mention that we found the matrix elements of the Gogny force in the $^{1}D_{2}$ channel to be very similar to those of $V_{low \, k}$. Consequently, the Gogny force does not provide any artificial pairing in that partial wave. Such a good agreement in the $D$-wave was not expected since the Gogny force was, as any phenomenological force in finite nuclei, adjusted without taking great care of its partial-wave content.

\begin{figure}
\begin{center}
\leavevmode
\centerline{\psfig{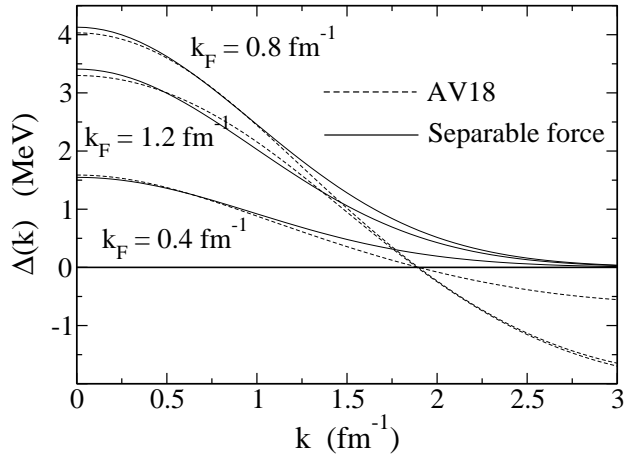}}
\end{center}
\caption{$\Delta_{k}$ obtained from $AV18$ and from the rank-one separable force for three different densities.}
\label{deltak}
\end{figure}

Due to the previous comparison, we have enough confidence in $V^{^{1}S_{0}}_{sep}$ to consider more detailed pairing properties. Beyond the ability of our force to reproduce $\Delta_{k_{F}}$, it is worth analysing the momentum dependence of the gap at fixed $k_F$. In Fig.~\ref{deltak}, the $\Delta_{k}$ obtained from $AV18$ and from our force are compared for three different densities~\cite{wiringa1}. The agreement is excellent in the energy and density intervals of interest ($k \leq 1.5 \, fm^{-1}, \, k_{F} \leq k^{sat}_{F} = 1.33 \, fm^{-1}$). The rank-one separable force is not designed to reproduce $\Delta_{k}$ for $k \geq 2 \, fm^{-1}$ where the $NN$ phase-shifts become negative. However, this is not a significant problem to describe pairing at low energy and density. Note that the momentum dependence of the gap is quite insensitive to the realistic interaction used, at least at low energy~\cite{baldo3,khodel}.

The good reproduction of $\Delta_{k}$ shows the ability of the force to describe fine pairing properties. It confirms that our choice for $v(k)$ is appropriate and that its range has been properly fitted. This comes back to the fact the gap $\Delta_{k} = \Delta_{0} \, v (k)$ is directly determined by the half on-shell $t$-matrix at $s=0$ ($\propto v (k)$), that is by the vertex function of the virtual state in the $^{1}S_{0}$ channel~\cite{brett2}.

\section{Effective pairing interaction at the mean-field level}
\label{Effectivepairing}

\subsection{Formalism}
\label{Formalism}

From the previous discussion, the property of the Gogny force as being close to a reduction of the bare force for low-energy phenomena can be understood. On the contrary, in spite of their success as phenomenological pairing interactions, DDDI cannot be interpreted as direct approximations of the bare force. One needs to understand them as effective vertices, which requires the derivation of an appropriate scheme. Also, our $V^{^{1}S_{0}}_{sep}$ bare force is still too complicated to be used in coordinate space HFB calculations. To deal with these two issues, we now recast the pairing problem in a slightly different manner. 

Let us start from the gap equation written at lowest-order in a given single-particle basis $\{a^{\dagger}_{i}\}$:

\begin{equation}
\Delta_{ij} \, = \, \sum_{mn} \, \langle \, i \, j \, |  \,  V \, | \, m \, n \, \rangle \, \kappa_{mn} \, \equiv \, \sum_{mnqr} \, \langle \, i \, j \, |  \,  V \, | \, m \, n \, \rangle \, F^{\, \Delta}_{mnqr} (0) \, \Delta_{qr}  \, \, \, ,
\label{gapequation3}
\end{equation}
where ${i,j,k,\ldots}$ are convenient labels to denote that single-particle basis and $V$ is the bare $NN$ force. We define $\Delta$ and $\kappa$ as the energy-independent pairing field and anti-symmetric pairing tensor respectively ($\kappa_{mn} = \langle \, \Phi \, |  \, a_{n} \, a_{m}  \, | \, \Phi \, \rangle$, where $| \, \Phi \, \rangle$ is the unperturbed quasi-particle HFB vacuum). For future convenience we have also introduced a two-body propagator $F^{\, \Delta}$ through $\kappa_{mn} \equiv \sum_{qr} F^{\, \Delta}_{mnqr} (0) \, \Delta_{qr}$.

Let us now introduce a two-body amplitude ${\cal R}$ in the medium through the equation\footnote{We make use of discrete sums in Eq.~\ref{2bodyscattmatrix}. Of course, an appropriate integral is to be considered when dealing with continuous quantum numbers. The specification to the $^{1}S_{0}$ channel of the interaction is implicit in the present work even if Eq.~\ref{2bodyscattmatrix} is valid in the general case.}:

\begin{equation}
\langle \, i \, j \, |  \,  {\cal R} (s) \, | \, k \, l \, \rangle \, = \, \langle \, i \, j \, |  \,  V \, | \,  k \, l \, \rangle \, + \, \sum_{mnqr} \, \langle \, i \, j \, |  \,  {\cal R} (s) \, | \, m \, n \, \rangle \, F^{\, {\cal R}}_{mnqr} (s) \, \langle \, q \, r \, |  \, V  \, | \, k \, l  \, \rangle \, \, \, ,
\label{2bodyscattmatrix}
\end{equation}
where $s$ is an external energy parameter and $F^{\, {\cal R}}$ a two-body propagator which will be specified later. Combining Eqs.~\ref{2bodyscattmatrix} and~\ref{gapequation3}, one can write:

\begin{equation}
\sum_{klqr} \langle \, i \, j \, |  \,  {\cal R} (s) \, | \, k \, l \, \rangle \, F^{\, {\cal R}}_{k \, l \, q \, r} (s) \, \Delta_{qr} \, = \, \sum_{klqrmnst} \, \langle \, i \, j \, |  \,  {\cal R} (s) \, | \, k \, l \, \rangle \, F^{\, {\cal R}}_{klqr} (s) \, \langle \, q \, r \, |  \,  V \, | \, s \, t \, \rangle \, F^{\, \Delta}_{stmn} (0)  \, \Delta_{mn} \, \, \, ,
\label{gapequation31}
\end{equation}
which, combined once more, allows recasting the gap equation under the form:

\begin{equation}
\Delta_{ij} \, = \, \sum_{mnqr} \, \langle \, i \, j \, |  \,  {\cal R} (s) \, | \, m \, n \, \rangle \, \left[F^{\, \Delta}_{mnqr} (0)  - F^{\, {\cal R}}_{mnqr} (s)\right]  \, \Delta_{qr} \, \, \, .
\label{gapequation4}
\end{equation}

No approximation has been done to derive Eq.~\ref{gapequation4} from Eq.~\ref{gapequation3}, and the derivation is valid whatever the two-body propagator $F^{\, {\cal R}}$ is. The solution of the pairing problem has to be the same by either using Eq.~\ref{gapequation3} or the set of coupled Eqs.~\ref{2bodyscattmatrix} and~\ref{gapequation4}. The previous rewriting of the gap equation is not restricted to the mean-field level. The propagators $F^{\, {\cal R}}$ and $F^{\, \Delta}$ may include off-shell nucleon propagation (fully or through a quasi-particle approximation) at finite temperature~\cite{bozek2,bozek1}. Eq.~\ref{gapequation4} still holds in that case. As we are interested in the mean-field treatment of the system at zero-temperature, we will only propagate the nucleons on-shell. Including dispersive effects in both the single-particle self-energy and the gap equation~\cite{bozek} constitutes a more advanced treatment of the many-body problem which is not conceivable nowadays for extensive calculations of finite nuclei. We have not tried to proceed to the same recast by starting from an irreducible vertex beyond the bare force. This could be done, in particular if using a quasi-particle approximation and/or considering a static limit for the higher-order terms in the interaction.

The ${\cal R}$-matrix is specified as soon as the choice of the two-body propagator $F^{\, {\cal R}}$ is made. Several choices are possible and selecting a particular one is simply a matter of convenience and of formal consistency. For several reasons discussed below, we require that the two-body effective vertex ${\cal R}$ treats particles and holes in a symmetric way and takes the superfluidity of the medium into account. More specifically, we define $F^{\, {\cal R}}$ in such a way that ${\cal R}$ sums through Eq.~\ref{2bodyscattmatrix} particle-particle and hole-hole ladders in the {\it presence} of pairing correlations. 

By considering the canonical basis associated with the actual Bogolyubov transformation solution of the problem and the hypothesis that time reversal symmetry is conserved\footnote{To be more specific, expressions~\ref{2bodypropmed} are rigorously valid only if the third part of the Bogolyubov transformation~\cite{bloch} is trivial.}, $F^{\, {\cal R}}_{mnqr} \, (s)$ and $F^{\, \Delta}_{mnqr} \, (s)$ read as~\cite{mehta1,henley,bozek}\footnote{Including off-shell propagation through a quasi-particle approximation would provide an additional factor $Z_{m}Z_{n}$, where $Z_{i}$ is the quasi-particle strength~\cite{bozek2}.}:

\begin{eqnarray}
F^{\, {\cal R}}_{mnqr} \, (s) \, &=& \, - \, \frac{(1-\rho_{m}) \, (1-\rho_{n})}{E_{m}+E_{n}+2s} \, \delta_{mq} \, \delta_{nr} \, \mp \, \frac{\rho_{m} \, \rho_{n}}{E_{m}+E_{n}+2s} \, \delta_{mq} \, \delta_{nr} \,  \, \, , \nb \\
&& \label{2bodypropmed} \\
F^{\, \Delta}_{mnqr} \, (s) \, &=& \, - \frac{1}{E_{m}+E_{n}+2s} \, \delta_{mq} \, \delta_{nr} \, \delta_{n\bar{m}} \, \, \, , \nb
\end{eqnarray}
where $\rho_{m} \equiv \rho_{mm} = [1-(\epsilon_{m}-\mu)/E_{m}] / 2 = \rho_{\bar{m}}$ embodies the (diagonal) one-body density matrix and $E_{m} = E_{\bar{m}} = \sqrt{(\epsilon_{m}-\mu)^{2}+\Delta^{2}_{m\bar{m}}}$ is a quasi-particle energy. The indices $(m,\bar{m})$ characterize the paired states in the canonical basis. The diagonal matrix elements of the single-particle mean-field $\epsilon_{m}$ involved in $E_{m}$ are defined using an appropriate scheme as already discussed in section~\ref{pairinggap}. Because of the hypothesis we have made concerning the Bogolyubov transformation, only the diagonal matrix elements $\Delta_{m\bar{m}}$ will be non-zero, selecting in Eqs.~\ref{gapequation4} and~\ref{2bodypropmed} the matrix elements of ${\cal R}$ and $F^{\, {\cal R}}$ involving paired two-body states of the same sort.

The first term entering the definition of $F^{\, {\cal R}}_{mnqr}$ in Eq.~\ref{2bodypropmed} sums particle-particle (p-p) ladders as in the $G$-matrix except for the fact that BCS occupation numbers and quasi-particle energies appear because of the superfluid nature of the system~\cite{mehta1}. The same is true for the second term delaing with hole-hole (h-h) ladders. Two different signs are considered to sum the diagrams associated with h-h ladders in ${\cal R}$. The ''$-$'' sign corresponds to what can be denoted as the ${\cal T}$-matrix in the superfluid system since ${\cal R}$ reduces in that case to the usual Galitskii $T$-matrix in the normal phase. This choice corresponds to summing the h-h ladders as they actually appear in the expansion of the ground-state energy when taking abnormal contractions into account in the theory~\cite{mehta1,henley}. Then, we will denote the in-medium matrix associated with the ''$+$'' sign as the ${\cal D}$-matrix. Note that even if the definition of this ${\cal D}$-matrix through Eq.~\ref{2bodyscattmatrix} is fully valid, it does not correspond to the summation of hole-hole ladder diagrams as they appear in the expansion of the ground-state energy of the system.

In the gap equation, it is appropriate to use ${\cal T/D}$ fully off-shell at $s=0$\footnote{In Eq.~\ref{2bodyscattmatrix}, a half or fully on-shell matrix element of ${\cal R}$ corresponds to $2s=E_{k}+E_{l}$. The choice $s=0$ corresponds to particular off-shell matrix-elements since it cannot be obtained by inserting any single-particle energies $\epsilon_{k}$ and $\epsilon_{l} \,$ into $2s=E_{k}+E_{l}$, as long as pairing correlations are present. Indeed, the quasi-particle energies are always different from zero in such a case.}. Eventually, Eq.~\ref{gapequation4} simplifies for the two considered reaction matrices into:

\begin{eqnarray}
\Delta_{i} \, \equiv \, \Delta_{i\bar{i}} \, &=& \, - \, \sum_{m} \, \langle \, i \, \bar{i} \, |  \,  {\cal D} (0) \, | \, m \, \bar{m} \, \rangle \,  2 \, \rho_{m} \, \frac{\Delta_{m}}{2 \, E_{m}} \nb \\
&& \label{gapequation5} \\
&=& \, - \, \sum_{m} \, \langle \, i \, \bar{i} \, |  \,  {\cal T} (0) \, | \, m \, \bar{m} \, \rangle \, 2 \, (1 - \rho_{m}) \, \rho_{m} \, \frac{\Delta_{m}}{2 \, E_{m}} \nb \, \, \, .
\end{eqnarray}

Again, different choices could have been made in Eq.~\ref{2bodyscattmatrix} for the two-body propagator, amounting to using other two-body vertices than the $\, {\cal T/D}$-matrices. For instance, choosing the plane-wave basis, reducing Eq.~\ref{2bodyscattmatrix} to the center of mass frame of the interacting pair and using Eq.~\ref{offshellt} for the two-body propagator, Eq.~\ref{2bodyscattmatrix} would reduce into the Lipmann-Schwinger equation defining the $t$-matrix. Our choice can in fact be seen as an extension to the superfluid medium of the regularizing procedures consisting of re-expressing the gap equation in terms of the scattering length~\cite{esbensen,marini,papenbrock} or of the $t$-matrix  in the vacuum~\cite{heiselberg4,khodel}. The latter choices are of particular interest in the low density regime where they allow the derivation of analytical formula for the gap in terms of known physical quantities. Eqs.~\ref{gapequation5} are not as useful in the low-density limit since the off-shell ${\cal T/D} (0)$ match $a^{^{1}S_0}$ only after the superfluity has disappeared.

However, taking ${\cal T/D} (0)$ is a formally optimal choice when studying nuclei at low energy. Such systems cover densities ranging from $0$ to $\rho_{sat}$, where the $S$-wave like-particle superfluidity evolves from the weak to the intermediate BCS regimes, before coming back to the weak coupling. Thus, it makes sense to use a two-body scattering matrix taking the varying density and superfluidity into account. Also, considering the in-medium vertex at the threshold $s=0$ makes the effective gap equation as given by Eq.~\ref{gapequation4} to have its simplest possible analytical forms (Eqs.~\ref{gapequation5}). The latter property is partly due to the fact the $\, {\cal T/D}$-matrices treat particles and holes on the same footing which is reasonable when dealing with pairing correlations. For that reason, it should be preferred to the Brueckner $G$-matrix~\cite{emery2}, even if calculated in the superfluid system~\cite{mehta1}. One could also have used the $T$-matrix in the normal phase. However, this vertex presents, like the $G$-matrix, a pole at the threshold ($\Leftrightarrow$ $s=\mu$ for the standard definition of its energy dependence) which signals the appearance of a Cooper bound state in the medium~\cite{emery2,thouless,bishop1,dickhoff}. As the aim of the gap equation is precisely to take care of the correlations associated with existing Cooper pairs, it should be combined with a regular vertex summing two-body correlations in the medium, except for those related to the possible bound-state. In fact, the Bogolyubov transformation aims to remove the pole from the scattering amplitude by defining a non-singular $\, {\cal T}$-matrix at the threshold ($s=0$ in the definition used above), while treating pairing correlations explicitly through the gap equation~\cite{mehta1,henley,bozek,balian}. 

An immediate by-product of the previous derivation is to show explicitly that none of the previously discussed in-medium matrices can be used in the gap equation without re-writing the latter accordingly~\cite{baldo3}. Aside from further aspects, the purpose of our recasting procedure is to incorporate the virtual high-energy transitions which appear in the original gap equation into the in-medium interaction. Whilst doing so, the important pair scattering around the Fermi surface are treated explicitly through Eq.~\ref{gapequation5}. This is reasonable since the gap equation is almost linear in the high momentum regime while it is highly non linear around $k_{F}$. Note that another recasting procedure was used in Refs.~\cite{baldo3,elgar2} to get stable solutions of the gap equation when solved with realistic bare $NN$ interactions. By summing virtual transitions above a sharp energy cut-off, an effective pairing interaction acting in a valence space was defined~\cite{baldo3,baldo0}. It was anticipated and shown that the microscopic effective force was close to the off-shell $T$-matrix at the threshold~\cite{baldo3,baldo1}. This result is not surprising in view of the previous discussion. 

The present scheme can now be translated into the definition of a microscopic effective pairing interaction to be used in the standard gap equation. By comparing Eqs.~\ref{gapequation3} and~\ref{gapequation5}, we introduce two versions of such an effective vertex whose matrix elements in the canonical basis read as:

\begin{eqnarray}
\langle \, i \, \bar{i} \, |  \,  V^{eff}_{{\cal D}} \, | \, m \, \bar{m} \, \rangle \, &\equiv& \, \langle \, i \, \bar{i} \, |  \,  {\cal D}^{^{1}S_0} (0) \, | \, m \, \bar{m} \, \rangle \, \, 2 \, \, \rho_{m} \, \, \, , \nb \\
&& \label{effectv} \\
\langle \, i \, \bar{i} \, |  \,  V^{eff}_{{\cal T}} \, | \, m \, \bar{m} \, \rangle \, &\equiv& \, \langle \, i \, \bar{i} \, |  \,  {\cal T}^{^{1}S_0} (0) \, | \, m \, \bar{m} \, \rangle \, \, 2 \, \, (1 - \rho_{m}) \, \rho_{m} \nb \, \, \, .
\end{eqnarray}

Each of these two versions includes a smooth cutoff as well as in-medium correlations. The cut-offs have appeared naturally when recasting the gap equation and will adapt self-consistently to the amount of pairing in the system. They do not have to be additionally chosen or optimized. They are measured with respect to $\mu$ and not with respect to the bottom of the mean-field potential and thus, evolve with density. Note that using ${\cal T}$ has led to the appearance of a symmetric cut-off on both sides of the Fermi energy while ${\cal D}$ comes together with a single cut-off above the Fermi energy. Again, one should stress that the effective interactions defined by Eq.~\ref{effectv} result from a rearrangement of the gap equation and that no approximation has been done in the mean-field treatment of pairing and that no effect beyond that level of approximation has been included at this point.

The two versions of $V^{eff}$ have been studied. In the following, we only present the results obtained by using the ${\cal D}$-matrix. The reason for that is because the ${\cal D}$-matrix is slightly less sensitive to the self-energy effects when those are included in the calculation as will be discussed in section~\ref{selfenergyeffects}. However, any important conclusion drawn in the following for the ${\cal D}$-matrix is valid for the ${\cal T}$-matrix.

\subsection{Infinite matter}
\label{Infinitematter}

\subsubsection{Calculation of the ${\cal D}$-matrix}
\label{caltmatrix}

Let us use the previous scheme to attack the infinite matter problem. The plane wave basis corresponds to the canonical basis associated with the Bogolyubov transformation solution of the problem. Using transparent notations, the two-body propagator involved in the equation of the ${\cal D}$ matrix reads as:

\begin{equation}
F^{\, {\cal D}}_{\vec{P}, \, \vec{k}} \, (s) \, = \, - \, \frac{1-\rho_{\vec{P}, \, \vec{k}}-\rho_{\vec{P}, -\vec{k}}}{E_{\vec{P}, \, \vec{k}}+E_{\vec{P}, -\vec{k}}+2s} \, .
\label{2bodypropinfmed}
\end{equation}

Performing the averaging over the angle $(\hat{\vec{P}},\hat{\vec{k}})$

\begin{equation}
\overline{F}^{\, {\cal D}}_{P, \, k} \, (s) \, = \, \frac{1}{4\pi} \, \int_{\phi=0}^{2\pi} \int_{\theta=0}^{\pi} \,  F^{\, {\cal D}}_{P, \, k, \, \cos \theta} \, (s) \, \sin \theta \, d\theta \, d\phi \, = \, \int_{x=0}^{1} \,dx \,   F^{\, {\cal D}}_{P, \, k, \, x} \, (s) \, \, \, ,
\label{angleaverage}
\end{equation}
and considering the separable form of $V^{^{1}S_{0}}_{sep}$, $\, {\cal D}^{^{1}S_0}$ can be integrated explicitly. Expressing ${\cal D}^{^{1}S_0}$ in the center of mass yields:

\begin{eqnarray}
\langle \, \vec{k} \, |  \,  {\cal D}^{^{1}S_{0}} (k_{F},P,s) \, | \, \vec{k} \, '  \, \rangle \, &=& \, \frac{\lambda \, v(k) \, v(k')}{1 - \lambda \, \int_{0}^{\infty} \frac{d \, k''}{2\pi^2} \, k'' \, ^2 \, v^2(k'') \, \overline{F}^{\, {\cal D}}_{P, \, k''} \, (s)} \, \label{solutioncaltmatrix} \\
&& \nb \\
&\equiv& \, \lambda \, \, v(k) \, \, h (k_{F},P,s) \, \, v(k') \, \, . \label{solutioncaltmatrixbis}
\end{eqnarray}

Starting from our separable force, Eq.~\ref{solutioncaltmatrix} shows that the in-medium vertex $\, {\cal D}^{^{1}S_{0}}$ is also separable in the three variables $(k, P, k')$. One has to check that using $\, {\cal D}^{^{1}S_{0}}$ in the gap equation as given by Eq.~\ref{gapequation5} provides the same gap as obtained in section~\ref{pairinggap}. Since only pairs with a zero total-momentum occur in nuclear matter and since the integration of $\, {\cal D}$ for $P=0$ does not require the angle averaging procedure, this reproduction must be exact. We have checked that it is so, both for the gap at the Fermi energy $\Delta_{k_{F}}$ and for $\Delta_{k}$. In the following, we want to study and {\it parametrize} the function $h$ characterizing the dependence of $\, {\cal D}^{^{1}S_{0}}$ on $k_{F}$ and $P$. This will be necessary to use $V^{eff}$ in HFB calculations of finite nuclei.

\subsubsection{Density dependence of $\, {\cal D}^{^{1}S_{0}} \, (0)$}
\label{depdens}

We now study the density dependence of $\, {\cal D}^{^{1}S_{0}} \, (0)$ in infinite matter. It enters the effective pairing force as a factor defined as $C(k_F) \equiv h (k_{F},0,0)$. 

\begin{figure}
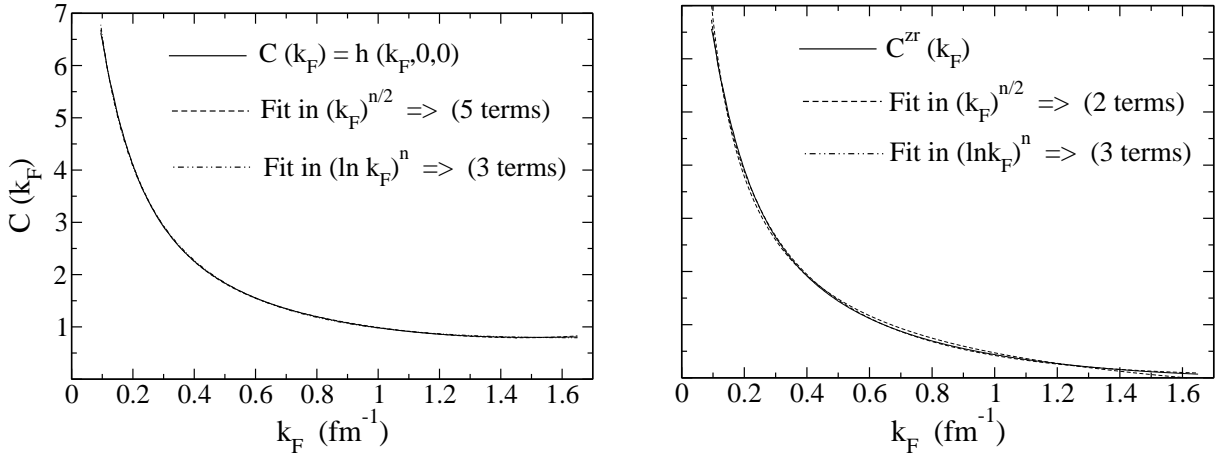

\begin{center}
\leavevmode
\centerline{\psfig{figure=cf_0.52_840.eps,height=6cm} \hspace{0.8cm} \psfig{figure=cf_0.0_840.eps,height=6cm}}
\end{center}
\caption{Left panel: derived density dependence of the effective pairing force in infinite matter (full line). Two different fits are also given. Right panel: same for the zero-range approximation of the effective pairing vertex.}
\label{cf1}
\end{figure}

This function, derived using $V^{^{1}S_{0}}_{sep}$, Eqs.~\ref{solutioncaltmatrix}-\ref{solutioncaltmatrixbis} and free single-particle energies, is shown in the left panel of Fig.~\ref{cf1}. The many-body effects in $\, {\cal D}^{^{1}S_{0}}$ are such that the magnitude of the in-medium interaction decreases with increasing density and saturates for $k_{F} \geq 1 \, fm^{-1}$. Within the usual framework of the local density approximation (LDA), this property translates into a pairing force which is slightly enhanced at the surface of the nucleus ($\Leftrightarrow k_{F} \approx 0.8-1.1 \, fm^{-1}$) as compared to the center ($\Leftrightarrow k^{sat}_{F} \approx 1.33 \, fm^{-1}$).

In infinite matter, the density dependence of the force has of course no link whatsoever with a surface effect. It is a pure density effect generated by the in-medium coupling of mean-field particles. It is only through the language of the LDA that one would talk about a {\it position-dependent} interaction in the nucleus. The present study clarifies this character of the effective interaction at the mean-field level. Additional effects, coming for instance from the induced interaction generated by the exchange of surface vibrations in finite nuclei, have to be considered on top of the mean-field approximation~\cite{barranco}.

\begin{figure}
\begin{center}
\leavevmode
\centerline{\psfig{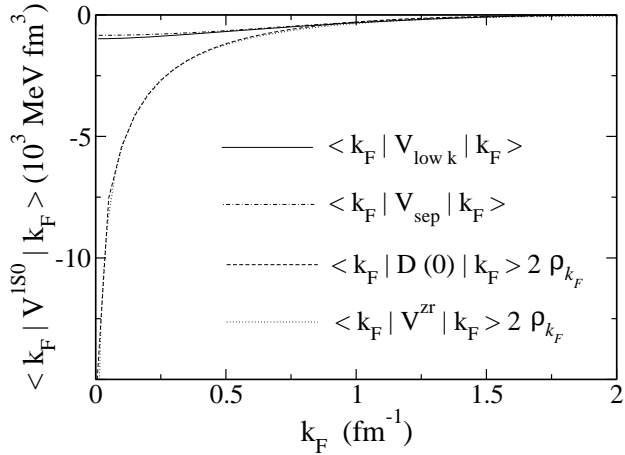}}
\end{center}
\caption{Diagonal matrix elements of different interactions at $k_{F}$ in the $^{1}S_0$ channel as a function of $k_{F}$ in infinite matter: $V_{low \, k}$ (full line), the separable bare force (dashed-dotted line), the finite-range effective pairing interaction (dashed line) and its zero-range approximation (dotted line).}
\label{analyseforces}
\end{figure}

The diagonal matrix elements of $V^{eff}_{{\cal D}}$ at $k_{F}$ are compared with those of $V_{low \, k}$ and of our separable force in Fig.~\ref{analyseforces}. Generally speaking, the pairing gap $\Delta_{k_{F}}$ is not determined by the diagonal matrix elements of the force at $k_{F}$ only~\cite{khodel}. Otherwise, the effective interaction, our separable bare force and AV18 could hardly give the same gaps. Obtaining the same gaps is possible because the different nature of their diagonal matrix-elements is compensated by different off-diagonal characters. For instance, the reduced  influence of the off-diagonal processes in the case of the effective force embodied by the cut-off $2 \, \rho_{k}$ is accompanied by an enhancement of its diagonal matrix elements through the density dependence. This shows why the in-medium effects resummed in the effective pairing force are correlated with the the cutoff emerging through the recast of the gap equation. Note that only in the case of such an effective interaction resumming off-diagonal processes (up to high energy if starting from a realistic bare $NN$ force) the gap is indeed very much determined by the matrix elements at $k_{F}$. This would be particularly true if using ${\cal T}$ instead of ${\cal D}$, since then the off-diagonal matrix elements are cut on both sides of the Fermi energy as shown by Eq.~\ref{gapequation5}. Only this property authorizes the use of weak coupling formulas~\cite{schulze2}.

\begin{figure}
\begin{center}
\leavevmode
\centerline{\psfig{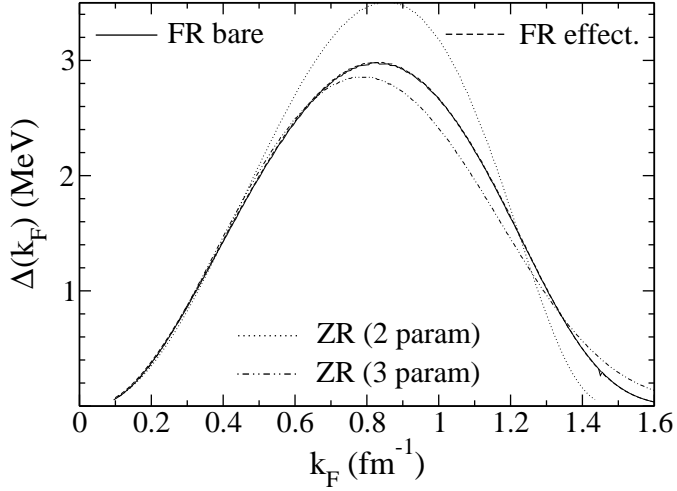}}
\end{center}
\caption{$^{1}S_{0}$ pairing gaps obtained from the bare force (full line), the effective force (dashed line) and its zero-range approximation with two different fits of its density dependence (dotted and dotted-dashed lines).}
\label{gapapprox}
\end{figure}

Later on, we will need $C (k_F)$ as an analytical function of $k_F$. For that purpose, we fit $C (k_F)$. One should stress that such a fitting procedure does not correspond to the use of free parameters in the definition of the pairing interaction. We simply reproduce $h (k_{F},0,0)$, derived with no freedom from $V^{^{1}S_{0}}_{sep}$, itself fitted on scattering properties and on the $^{1}S_{0}$ pairing gap provided by $AV18$. As the gap is exponentially sensitive to the strength of the interaction, a fine fit of $C (k_{F})$ is required to reproduce $\Delta_{k_{F}}$ with high accuracy. Two different fits are compared with the exact function $C (k_{F})$ on the left panel of Fig.~\ref{cf1}. The gap re-calculated using the fit defined through powers of $\ln k_{F}$ is compared with the gap derived from the bare force on Fig.~\ref{gapapprox}. They are nearly identical.

The two fits displayed on Fig.~\ref{cf1} have a quite different analytical character. Using enough terms, one can actually propose a large set of precise fits making use of very different power series. For a given number of terms, some expansions are more precise than others. For instance, an expansion in powers of $k_F$ requires a large number of terms whereas expansions in powers of $\sqrt k_F$ or $k_F^{-1/4}\ldots$ converge much faster. Whatever the chosen expansion, the important feature is to reproduce precisely the behavior of $C (k_{F})$ derived from $V^{^{1}S_0}$. 

The fit making use of powers of $\ln k_{F}$ provides an excellent approximation of $C (k_{F})$ with only three terms and presents the particular feature that adding other powers of $\ln k_F$ does not improve the fit significantly. As this was not the case for any other expansion we have tried, we consider this function as an ''exact'' analytical form of $C (k_{F})$. This function rises in the limit $k_{F} \rightarrow 0$. This agrees with the fact that ${\cal D}^{^{1}S_0} (0)$ tends to the off-shell $t^{^{1}S_0}$-matrix at $s=0$, which almost diverges because of the virtual state in the vacuum at approximately zero scattering energy. This is seen in Fig.~\ref{analyseforces}. As no real bound state exists in this channel, the pairing collapses accordingly in the $k_{F} \rightarrow 0$ limit as seen in Fig.~\ref{gapapprox}.

\subsubsection{Zero-range approximation}
\label{zerorange}

An appealing property of the scheme developed here is its suitability for studying a zero-range approximation of the effective pairing interaction. Since the high-energy transitions appearing in the gap equation have been resummed into the in-medium vertex $\, {\cal D}^{^{1}S_{0}}$, no divergence is expected when taking its zero-range limit. This is embodied by the presence of the cut-off $2 \, \rho_{k}$ in the gap equation.

Expanding $\langle \, \vec{k} \, |  \,  {\cal D}^{^{1}S_{0}} (k_{F},P,0) \, | \, \vec{k} \, '  \, \rangle$ in the range and in the non-locality leads to an effective pairing interaction reading in momentum space as:

\begin{equation}
\langle \, \vec{k} \, |  \,   V^{zr} (k_{F}) \, | \, \vec{k} \, '  \, \rangle \, \approx \, 2 \, \, \lambda \, \, C^{zr} (k_{F}) \, \, \rho_{k'} \, \, \, .
\label{solutioncaltmatrixter}
\end{equation}

One is simply left with a density-dependent force, constant in $k$ and varying as $\rho_{k'}$, where $k'$ is the momentum of the intermediate state in the gap equation. The local approximation consisting of taking $V^{zr}$ independent of $P$ is shown to be appropriate in section~\ref{totmomentumdenp}. The gap equation in infinite matter associated with the force defined through Eq.~\ref{solutioncaltmatrixter} provides a constant solution $\Delta = \Delta^{zr}_{k_{F}}$. 

The structure of the zero-range approximation of the effective pairing force we obtain here provides a formal justification for the use of DDDI complemented with a smooth cut-off above the Fermi energy~\cite{bonche1}. Considering the ${\cal T}$-matrix instead of the ${\cal D}$-matrix would allow a similar justification for those DDDI which have been used with a symmetric cut-off on each side of the Fermi energy~\cite{rigol}.

We now extract the function $C^{zr} (k_{F})$. This is done by matching $\Delta^{zr}_{k_{F}}$ with $\Delta_{k_{F}}$ obtained from AV18 as a function of $k_{F}$. The function $C^{zr} (k_{F})$ is plotted on the right panel of Fig.~\ref{cf1}. As expected, no difference is seen at very low density between $C (k_{F})$ and $C^{zr} (k_{F})$, whereas the effect of the finite range becomes more and more important with increasing density. While $C (k_{F})$ was almost flat and non-zero when approaching the saturation density of nuclear matter, the zero-range in-medium pairing interaction goes to zero. In other words, the present calculation {\it predicts} (through the LDA) that, when approximated by a zero-range-like vertex, the effective pairing interaction has to be renormalized by a density-dependent intensity whose surface character is more pronounced than in the case of the finite-range original version. Quantitatively speaking, while $C (k_{F})$ is multiplied by a factor 1.4 between $k^{sat}_{F} = 1.33 \, fm^{-1}$ and $k_{F} = 0.8 \, fm^{-1}$ ($\approx \rho_{sat} / 5$), $C^{zr} (k_{F})$ is multiplied by a factor 3.6. The zero-range vertex has a behavior between surface and volume. This confirms the results obtained from refined phenomenological studies performed with these kinds of vertices~\cite{duguet2,doba4,doba2,doba7}. 

The different behaviors of the finite-range and zero-range forces can be understood in the following way. The density dependence saturates when the size of the Fermi sea is of the order of the inverse of the interaction range. Beyond that point the range governs the coupling inside the Fermi sea. This is visible on the left panel of Fig.~\ref{cf1} where the finite-range interaction saturates beyond $k_{F} \geq 1/\sqrt\alpha \approx 1.4 \, fm^{-1}$). For the zero-range vertex the same only happens when $k_{F} \rightarrow 1/\sqrt\alpha \equiv \infty$ where the interaction goes to zero in order to compensate for its artificial constant coupling in $k$-space.

It becomes clear in the present context that the surface enhancement of a zero-range pairing vertex takes care to some extent of the finite range of the nucleon-nucleon interaction. The effect of the range, although noticeable, seems to be re-normalizable at the mean-field level. This is seen on Fig.\ref{analyseforces} where the diagonal matrix elements of the two interactions at the Fermi surface are compared. They are nearly identical. The possible renormalization of the range conveys that nuclear matter presents a so-called weak-coupling BCS regime over the density range of interest characterized by the size of the Cooper pairs being much larger than the inter-particle distance and the range of the force. Such a result provides the grounds to the local density functional theory~\cite{kohn1,kohn2} in the context of superfluid nuclear matter~\cite{bulgac1,bulgac2}, even if the context is slightly different in that case since all beyond-mean-field effects are included in the in-medium coupling constants. 

However, the fact that the range and the non-locality of the force do not need to be resolved has to be confirmed in finite nuclei. Indeed, a zero-range vertex carries less information than the finite-range force does; this being embodied in the present case by the different non-diagonal matrix elements of our finite- and zero-range forces. While the zero-range version is able to reproduce the gap at the Fermi level as a function of the density, it predicts constant gaps as a function of momentum at a given density. Such an approximation is doubtful, especially around the Fermi energy where the gap is rapidly varying~\cite{baldo3}, as seen in Fig.~\ref{deltak}. Treating the range and the non locality of the force affects the state-dependent pairing gaps, the particle-width of deep-hole states in finite nuclei and translates to some extent into the spatial character of the pairing field~\cite{doba8}. It may also be of importance to describe excited states in nuclei. Resolving this issue requires an excursion far from the valley of $\beta$ stability~\cite{doba8}.

Using $C^{zr} (k_{F})$ within the LDA requires the identification of its analytical dependence on $k_{F}$. Unlike the function $C (k_{F})$, derived from the bare force through the calculation of the ${\cal D}$-matrix, fitting $C^{zr} (k_{F})$ amounts to fixing the free parameters entering the definition of the phenomenological zero-range vertex. Two examples are plotted on the right panel of Fig.~\ref{cf1}. The overall multiplication by $\lambda$ in Eq.~\ref{solutioncaltmatrixter} does not correspond to adding another free parameter. No parameter is needed to specify the cut-off. 

Fig.~\ref{gapapprox} displays the gap re-calculated by inserting the two used fits into Eq.~\ref{solutioncaltmatrixter}. The three-parameter fit allows a satisfactory reproduction of the gaps derived from the finite-range interaction. The fit employing two parameters does not provide sufficiently precise results. This last parametrization overshoots the gap around its maximum by half an MeV and undershoots it beyond $k_{F} = 1.25 \, fm^{-1}$. 

The number of parameters usually used in connection with zero-range forces is two for a pure-volume pairing and three for a density-dependent pairing~\cite{bertsch,rigol,garrido1,garrido2,doba1}. The reproduction of the gap in nuclear matter~\cite{garrido1,garrido2} as well as recent calculations in finite nuclei~\cite{doba1,duguet2} favor a density dependence. In such studies, the exponent of the density has sometimes been varied, in which case it could be considered as an additional free parameter. Even if the presently derived zero-range force also makes use of three parameters, it corresponds to a cleaner parametrization since the cut-off appeared naturally and the adjustment of the force is simply a matter of reproducing a fixed, derived function.

Following the path of a range expansion, one could include correction terms in $k^2, k' \, ^2, k^2*k' \, ^2 \ldots$ and obtain the corresponding density dependence by using the previous method. Finally, it is worth noting that the finite-range effective interaction makes use of only two parameters for an even better quality of results.

\subsubsection{On-shell self-energy and isospin dependence of $\, {\cal D}^{^{1}S_{0}} \, (0)$}
\label{selfenergyeffects}

The logarithmic density dependence of the effective interaction has been derived from the integration of the $\, {\cal D}$-matrix to {\it all orders} in the bare interaction. This integration includes both hole-hole and particle-particle ladders in the superfluid phase and treats the energy denominator of the two-body propagator explicitly. However, while everything has been consistently compared so far using free single-particle energies, the parametrizations of the functions $C (k_{F})$ and $C^{zr} (k_{F})$ obtained without including self-energy effects are of course approximate. As mean-field calculations of finite nuclei will eventually be performed, the function $C (k_{F})$ should be derived accordingly. Consequently, we recalculate the effective interaction including on-shell self-energy effects through a density-dependent effective-mass approximation for the single-particle energies:

\begin{equation}
\epsilon_{k_{F},\beta} (k) \, = \, \frac{k^{2}}{2 m^{\ast} (k_{F},\beta)} \, + \, \epsilon_{k_{F},\beta} (0) \, \, \, ,
 \label{effectmassapprox}
\end{equation}
where $k_{F}^3 = 3\pi^2\rho/2 = [(k^n_{F})^3 + (k^p_{F})^3]/2$ and $\beta=(\rho_{n}-\rho_{p})/\rho = [(k^n_{F})^3-(k^p_{F})^3]/2k_{F}^3$ are the Fermi momentum and the matter asymmetry, respectively. Quantities relating to a particular isospin value present an index $q$ either specified as $n$ and $p$ or by $1/2$ and $-1/2$ when dealing with neutron and proton, respectively. 

\begin{figure}
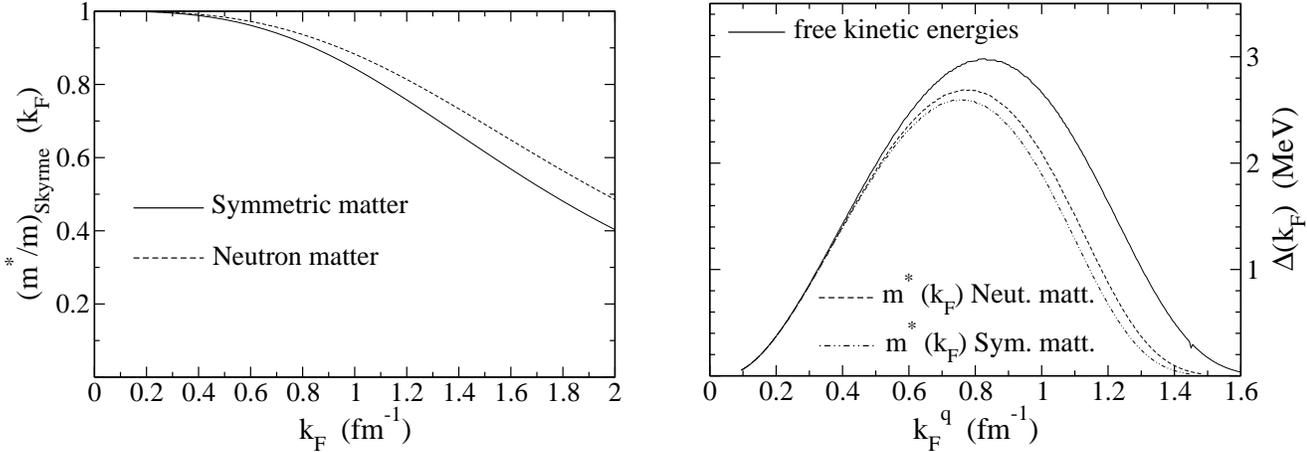

\begin{center}
\leavevmode
\centerline{\psfig{figure=effectmassSkyrme.eps,height=6cm} \hspace{0.8cm} \psfig{figure=gapeffectmass.eps,height=6cm}}
\end{center}
\caption{Left panel: neutron effective mass in neutron (dashed line) and symmetric (full line) matter for the SLy4 Skyrme force. Right panel: $^{1}S_{0}$ gaps calculated with three choices of single-particle energies: free kinetic energies (full line), effective mass approximation in neutron matter (dashed line) and effective mass approximation in symmetric matter (dotted-dashed line).}
\label{effectmasssk}
\end{figure}

We use the effective mass as given by a standard parametrization of the Skyrme force~\cite{chab}. Such an effective mass mimics an average of the Brueckner-Hartree-Fock (BHF) $k$-mass:

\begin{equation}
\frac{m^{\ast}}{m} \, (k, k_{F},\beta) \, = \, \left( \frac{m}{k} \, \frac{\partial \, \epsilon^{\, BHF}_{k_{F},\beta} (k)}{\partial \,k}\right)^{-1} \, \, \, ,
 \label{effectmassBHF}
\end{equation}
over the Fermi sea. It is independent of $k$, smaller than one at all densities and of the order of $0.7$ at saturation density of symmetric nuclear matter~\cite{schulze}. The Skyrme effective-masses in symmetric and neutron matter are plotted on the left panel of Fig.~\ref{effectmasssk} for the SLy4 parametrization~\cite{chab}. As seen, self-energy effects are larger in symmetric matter than in neutron matter at a given density because of the stronger neutron-proton interaction.

As the gap equation is particularly sensitive to the density of states at the Fermi surface, the use of an averaged $k$-mass which does not reflect the bump at the Fermi surface of the actual BHF $k$-mass is questionable. When dealing with pairing, one should maybe consider the BHF $k$-mass at $k_{F}$ instead~\cite{baldo4}, or the complete BHF single-particle energies. However, two arguments are in favor of the averaged effective masses. First, the bump of the $k$-mass at $k_{F}$ is smoothed out to some extent by the presence of pairing correlations~\cite{lombardo2}. Second, the main purpose of the present section is to include self-energy effects in  $\, {\cal D}^{^{1}S_{0}}$ which should not be as sensitive as the gap to the density of states at $k_{F}$.

The corresponding gaps for symmetric and neutron matter are compared on the right panel of Fig.~\ref{effectmasssk} to the one obtained using free kinetic energies. The gaps are plotted as a function of the Fermi momentum of the species (neutrons or protons) concerned by the pairing. As expected, the inclusion of on-shell self-energies reduces the gap. However, the effective mass obtained from the SLy4 parametrization is such that the reduction of the gap in neutron matter is too strong compared to the one obtained using realistic bare interactions and BHF single-particle energies~\cite{baldo3,zuo}. On the other hand, the reduction is slightly too small in symmetric matter, especially at low density~\cite{baldo3,zuo}. This corresponds to isoscalar and isovector parts of the effective mass which are respectively too large and too small. In any case, the global effect is present at the densities of interest. This should be sufficient to discuss the effects of the self-energy on the effective pairing interaction.

The modified functions $C (k_{F}^{q})$ and $C^{zr} (k_{F}^{q})$ have been re-calculated for neutron and symmetric matter. They are nearly identical to those derived using free kinetic energies, at least up to saturation density. Therefore, one can safely state that the vertex only depends on the density of the interacting nucleons $\rho_{q} = k_{F}^{q}/3\pi^{2} = \rho \, [1+(-1)^{1/2-q} \, \beta]/2$ and is not influenced by the surrounding nucleons of the other species. This result offers a microscopic answer to the issue dealing with the isovector density dependence of the effective pairing force at the mean-field level. A dependence on the total density as used so far is not justified from the microscopic point of view. This could of course change beyond the mean-field~\cite{heiselberg4}. 

At the present stage, the best logarithmic fit of the density dependence of the finite range effective pairing interaction is given when using three terms by:

\begin{equation}
C (k_{F}) \, \approx \, 0.978444 - 0.682204 \ln k_F + 0.761575 \, (\ln k_F)^{2} \, \, \, \, ,
\label{fit2cf}
\end{equation}
while in the case of its zero-range approximation, we find:

\begin{equation}
C^{zr} (k_{F}) \, \approx \, 0.420637 - 1.012900 \ln k_F + 0.708922 \, (\ln k_F)^{2} \, \, \, \, .
 \label{fit3cf}
\end{equation}
 
Using these fits with the Skyrme effective mass approximation, the gaps obtained in neutron and symmetric matter reproduce well those shown in Fig.~\ref{effectmasssk}. Finally, note that the energy functional of normal and abnormal densities obtained with such density-dependent vertices in the particle-particle $(T=1, T_{z}=\pm1)$ channels is isospin symmetric.

\subsubsection{Total-momentum dependence of $\, {\cal D}^{^{1}S_{0}} \, (0)$}
\label{totmomentumdenp}

We now study the dependence of the ${\cal D}$-matrix on the total momentum $P$ of the interacting nucleons. As nucleons with non-zero total momenta can be paired in finite systems, it is important to get insights into the corresponding component of the effective force. In order to obtain the ${\cal D}$-matrix for $P \neq 0$, a numerical averaging over the angle between the relative and total momenta of the pair in the intermediate states has been performed in Eq.~\ref{angleaverage}.

\begin{figure}
\begin{center}
\leavevmode
\centerline{\psfig{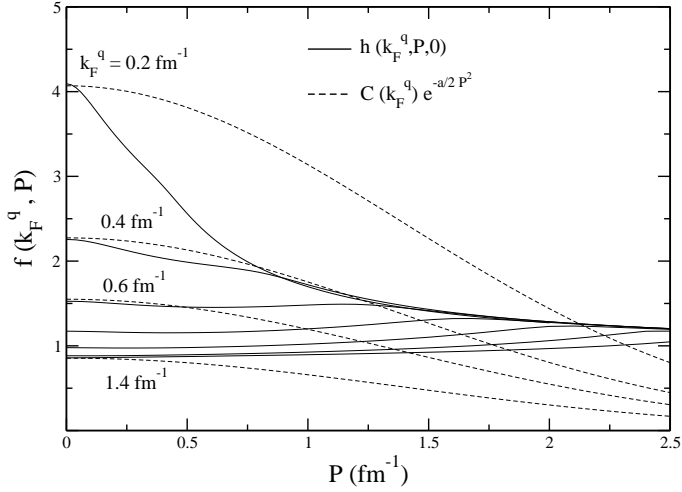}}
\end{center}
\caption{Total momentum dependence of the ${\cal D}^{^{1}S_{0}}$-matrix for several densities (full lines). A comparison is done with the function $C (k_{F}^{q}) \,e^{-\alpha^{2} P^{2}/2}$ (dashed lines).}
\label{depP}
\end{figure}

The function $h(k_{F}^{q},P,0)$ is plotted on Fig.~\ref{depP} as a function of $P$, up to $P= 2.5 \,  fm^{-1}$, for densities ranging from $k^{q}_{F} = 0.2 \, fm^{-1}$ to $k^{q}_{F} = 1.4 \, fm^{-1}$. The value $P= 2.5 \,  fm^{-1}$ corresponds to the maximum $P$ a pair of nucleons inside the Fermi sea can have at $k^{q}_{F} = 1.25 \, fm^{-1}$. The $P$ dependence of ${\cal D}$ is significant at extremely low density, whereas it becomes less pronounced with increasing $k^{q}_{F}$. Around saturation density, the interaction is almost independent of $P$. Also, the interaction is strongly modified at low $P$ when increasing the density while it is less sensitive to the medium for a pair having a large total momentum.

For a reason which will be clarified later, we need to test the hypothesis that $h (k_{F},P,0)$ is separable in $(k_{F},P)$ and that the total momentum dependence can be parametrized by $f (P) \equiv e^{-\alpha^{2} P^{2}/2}$, $\alpha$ being the same range as before. As a matter of comparison, the product $C (k_{F}^{q}) \, e^{-\alpha^{2} P^{2}/2}$ and the function $h(k_{F}^{q},P,0)$ are plotted on Fig.~\ref{depP}. The agreement is not satisfying at extremely low density where $h$ decreases much faster than $C (k_{F}^{q}) \, f (P)$ as a function of $P$. While the parametrization works well for typical surface densities, it decreases too quickly as a function of $P$ around saturation density. Of course, a pair of nucleons formed in finite nuclei has non-zero components for all values of $P$. In addition, the intensity of the interaction correlating the pair for each of these components must be seen as an average over all densities. As a result, we expect to have a good description of the combined density and total-momentum dependences of the effective pairing force by parametrizing $h(k_{F}^{q},P,0)$ through $C (k_{F}^{q}) \, f (P)$.

\subsection{Effective pairing interaction in coordinate space}
\label{interactionrealspace}

To obtain the previous results in infinite matter, we started naturally from the bare force expressed in the plane-wave basis and derived the effective pairing interaction accordingly. To perform calculations in finite nuclei, and especially if solving the problem in coordinate space, it is necessary to have the expression of the interaction as a function of the interacting nucleon positions. As already mentioned, the force presently used is finite ranged and non local. Consequently, it depends on the four position vectors $\{\vec{r}_{1},\vec{r}_{2},\vec{r}_{3},\vec{r}_{4}\}$ of the incoming and outgoing interacting nucleons. Also, the effective nature of the force has been characterized by its density dependence $C (k_{F}^{q})$. As the spin-isospin part of the force is trivial, let us consider the spatial part of the two-body interaction under the form:

\begin{equation}
\langle \, \vec{r}_{1} \, \vec{r}_{2} \, |  \,  {\cal D}^{^{1}S_0}_{q} (0) \, | \, \vec{r}_{3} \, \vec{r}_{4} \, \rangle \, = \, \frac{\lambda}{(2\pi)^{6}\alpha^{12}} \, \int \, d \vec{r} \, \, C \left(\rho_{q}(\vec{r} \,)\right) \, \, e^{-\sum_{i=1}^{4} \, |\vec{r}-\vec{r}_{i}|^{2} / 2 \alpha^{2}} \, \, \, \, \,,
\label{2bodyrealspace1}
\end{equation}
where 

\begin{equation}
C \left(\rho_{q}(\vec{r} \,)\right) \, = \, 1.179341 + 0.345992 \, \ln \rho_{q}(\vec{r} \,) + 0.084619 \, \left[\ln \rho_{q}(\vec{r} \,)\right]^{2} \, \, \, \, \, \, ,
\label{formfactor}
\end{equation}
is the form factor resulting from the local-density approximation of the function $C (k_{F}^{q})$ derived in infinite matter and discussed in section~\ref{depdens}. The fit expressed through Eq.~\ref{fit2cf} and the relationship $k_{F}^{q}=(3\pi^{2}\rho_{q})^{1/3}$ has been used to write Eq.~\ref{formfactor}.

The form given by Eq.~\ref{2bodyrealspace1} is reasonable since, if one forgets about the density dependence (or takes it to be constant in space), the following identity holds:

\begin{equation}
\langle \, \vec{r}_{1} \, \vec{r}_{2} \, |  \,  {\cal D}^{^{1}S_0}_{q} (0) \, | \, \vec{r}_{3} \, \vec{r}_{4} \, \rangle \, = \, \frac{\lambda}{(2^{5}\pi^{3})^{3/2}\alpha^{9}} \, e^{-\sum_{i<j}^{4} \, |\vec{r}_{i}-\vec{r}_{j}|^{2} / 8 \alpha^{2}} \, \, \, \, \, .
 \label{2bodyrealspace2}
\end{equation}

Hence the interaction simply generalizes the Gaussian form to the non-local case. The matrix elements of the interaction defined by Eq.~\ref{2bodyrealspace1} calculated in the plane-wave basis take the form:

\begin{equation}
\langle \, \vec{k}_{1} \, \vec{k}_{2} \, |  \,  {\cal D}^{^{1}S_0}_{q} (0) \, | \, \vec{k}_{3} \, \vec{k}_{4} \, \rangle \, = \, (2\pi)^{3} \, \, \lambda \, \, C (k_{F}^{q}) \, e^{-\alpha^{2} \, \left(k^{2}   \, +  \, \frac{P^{2}}{2}  \, +  \, k^{' \,2}\right)}  \, \, \delta (\vec{P}-\vec{P} \,') \, \, \, \, \, .
 \label{2bodyrealspace3}
\end{equation}

This is precisely the ${\cal D}^{^{1}S_0}_{q}$-matrix derived in infinite matter from our separable bare force and studied in previous sections. To reach such a form, one single approximation dealing with the $P$ dependence of ${\cal D}^{^{1}S_0}_{q}$ was performed in section~\ref{Infinitematter}. Thus, Eq.~\ref{2bodyrealspace3} clarifies why the approximate dependence on $P$ was needed in order to write the effective pairing interaction in coordinate space. Note that the parameters of the force have been fixed in infinite matter and that no room is left for any adjustment in finite systems.

It is worth characterizing the physical content of the LDA used for the function $C (k_{F}^{q})$ in Eq.~\ref{2bodyrealspace1}. The particle-particle and hole-hole ladders (including all high energy processes) associated with two-body {\it scattering} in the medium are resummed in the effective interaction by considering the nucleus as slices of homogeneous, infinite nuclear matter. However, the scattering of pairs around the Fermi surface responsible for the formation of the {\it bound states} in the medium are treated explicitly in the finite system through the resolution of the HFB equations. Also, the quantum shell effects which strongly influence the latter pair scatterings are taken into account explicitly. All along the way, the finite range and the non-locality of the interaction is fully considered. We expect the local approximation for the correlations associated with two-body scattering to satisfactory and the inclusion of gradient terms\footnote{This is of course to be differentiated from gradient terms simulating beyond-mean-field effects like the exchange of surface modes.} to improve on it to be rather unnecessary~\cite{delion}. The fact that the LDA is performed on the effective vertex itself makes the present approximation different from a so-called semi-classical treatment~\cite{kucharek} or from the local density functional theory~\cite{kohn1,kohn2,bulgac1,bulgac2}. In particular, it has allowed us to derive fine details of the interaction, such as its non-analytical low-density or isovector character, which are otherwize difficult to identify by looking directly at total-energy differences because of our presently limited experimental knowledge~\cite{bulgac3}.

One of the most important properties of the newly defined pairing interaction can be identified by calculating its antisymmetrized matrix elements in any given single-particle basis. To be explicit, we define a basis $\{\varphi_{np\zeta q}\}$ where $(n,p,\zeta,q)$ denote the principal quantum number, the parity, the z-signature and the isospin of the state, respectively. This is typical of the HF or canonical basis of a (triaxial, rotating, odd) deformed nucleus. Specifying $\varphi_{np\zeta qs}$ as the component of $\varphi$ having a good projection $s$ of the spin on the z-axis (deformation axis in the intrinsic frame), the antisymmetrized matrix elements of the effective pairing force as defined through Eqs.\ref{effectv} and \ref{2bodyrealspace1} read as:

\begin{eqnarray}
\overline{(V^{eff}_{q})}_{ikjl} \, &=& \,  \lambda \, \left(\rho_{j} + \rho_{l}\right) \, \sum_{ss'} \, \int \, d \vec{r} \, \, C \left(\rho_{q}(\vec{r} \,)\right) \, \, \tilde{\varphi}^{\, \ast}_{n_ip_i\zeta_i qs} (\vec{r} \,) \, \tilde{\varphi}^{\, \ast}_{n_kp_k\zeta_k qs'} (\vec{r} \,) \label{2bodyrealspace44} \\
&& \hspace{2cm} \times \, \left\{\tilde{\varphi}_{n_jp_j\zeta_j qs} (\vec{r} \,) \, \tilde{\varphi}_{n_lp_l\zeta_l qs'} (\vec{r} \,) - \tilde{\varphi}_{n_jp_j\zeta_j qs'} (\vec{r} \,) \, \tilde{\varphi}_{n_lp_l\zeta_l qs} (\vec{r} \,) \right\} \, \, \, \,  , \nb
\end{eqnarray}
where $\tilde{\varphi}_{np\zeta qs}$ is defined as:

\begin{equation}
\tilde{\varphi}_{np\zeta qs} (\vec{r} \,) \, = \, \frac{1}{(\sqrt{2\pi} \, \alpha)^{3}} \, \int \, d \vec{r} \, ' \, e^{-|\vec{r}-\vec{r} \, '|^{2} / 2 \alpha^{2}} \, \varphi_{np\zeta qs} (\vec{r} \, ') \, \, \, \,  .
 \label{2bodyrealspace5}
\end{equation}

Restricting the pairing to states of opposite signatures in the Bogolyubov transformation provides the matrix element~\ref{2bodyrealspace44} with an additional factor $\delta_{-\zeta_i\zeta_k} \, \delta_{-\zeta_j\zeta_l}$. Writing the matrix elements in the HF basis requires the use of the cut-off $(\rho_{j} + \rho_{l})$ instead of $2 \, \rho_{m}$ in the canonical basis. This is a natural, but not fully rigorous extension as can be realized by going back to the recasting procedure proposed in section~\ref{Formalism}. One should preferably work in the canonical basis.

Solving the HFB problem in coordinate space is a nice feature since it allows a natural treatment of all kinds of deformations and is well suited to describe exotic systems for which asymptotic properties of individual wave-functions and densities must be considered carefully. However, the naive use of a finite-range interaction in this context is numerically prohibitive~\cite{ring1,doba6,grasso2}. Indeed, solving the BCS gap equation is too costly for practical applications, while the HFB problem takes the form of a coupled set of integro-differential equations, also untractable for systematic studies. This is the main reason why zero-range forces have been extensively used so far~\cite{bertsch,terasaki1}. The matrix elements of our finite-range, non-local effective pairing-interaction look very similar to those obtained from a zero-range force~\cite{terasaki4}. The only additional cost is to replace the single-particle wave-functions $\varphi_{np\zeta qs}$ by their convoluted counterpart as defined through Eq.~\ref{2bodyrealspace5}. This property makes the corresponding HFB problem in (3-dimensions) coordinate space tractable through the two-basis method~\cite{gall,terasaki4,yamagami}, and almost equivalent computationally to the use of a local, zero-range force. In fact, only trivial modifications of HFB codes using this method are required. Adding a routine to convolute the HF wave-functions, using the proposed density-dependence and inserting the factor $(\rho_{j} + \rho_{l})$ when calculating the matrix elements of the pairing field in the HF basis is necessary. Of course, the pairing field provided by our force is non local. This feature prevents from solving the HFB problem through a direct diagonalization of the HFB matrix in coordinate space~\cite{doba6,teran}.

It is also worth noting that the derivative of the pairing energy with respect to the density matrix $\{\rho_{ji}\}$ when minimizing the total energy to obtain the HFB equations by must not be considered with the present force. Indeed, our effective pairing interaction is equivalent to the bare force in the pairing channel and its dependence on $\{\rho_{ji}\}$ simply arises through a recasting of the gap equation originally written in terms of the bare force. Of course, this statement is no longer valid when going to higher orders, when renormalizing the effect of the three-body force~\cite{fayans2} or when using the philosophy of the density functional theory.

As an alternative to our finite-range and non-local microscopic effective vertex, one can use its zero-range approximation studied in section~\ref{zerorange}. When performing the zero-range limit appropriately, the convolution used in Eq.~\ref{2bodyrealspace5} becomes the identity operation. Consequently, the matrix elements of the zero-range vertex take the form given by Eq.~\ref{2bodyrealspace44} with $\tilde{\varphi}_{n_lp_l\zeta_l qs}$ replaced by $\varphi_{n_lp_l\zeta_l qs}$ and $C \left(\rho_{q}(\vec{r})\right)$ replaced by the LDA of the function $C^{zr} (k_{F}^{q})$ parametrized accordingly.

Last but not least, it is essential to note that the separable bare force defined through Eq.~\ref{approxmatrixelement} does not lead to simple calculations in coordinate space. Its finite range and particular form of non-locality make it numerically untractable in such a context. This is only by going to the in-medium vertex that one obtains an interaction as given by Eq.~\ref{2bodyrealspace1}.

\section{Discussion}
\label{Discussion}

While very low-densities are not essential for stable nuclei, they become increasingly important when going to exotic systems. Indeed, nuclei close to the neutron drip-line develop extended low-density halos or skins where pairing correlations play a crucial role. This is also true in neutron stars' crust. Thus, the behavior of the pairing force at low density is of great interest for the study of these systems. It is clear that usual DDDI miss the very low-density part of $C^{zr} (k_{F})$ discussed in section~\ref{zerorange}. In Ref.~\cite{doba2}, strongly increasing interactions at low density were simulated through small exponents $\gamma$ in Eq.~\ref{DDDI}. Such pairing interactions were disregarded, notably because of the unrealistic reduction of the two-neutron separation energy across the magic number $N=82$. However, these interactions were used together with a phenomenological cut-off at a maximal fixed energy in the single-particle spectrum. In the present case, the cut-off $2 \, \rho_{m}$ derived in connection with the density dependence will weight the low-density content of the force in a very different way. 

Also, the isovector character of the pairing force should manifest itself when studying drip-lines systems. The dependence of our force on the density of the interacting nucleons will provide a weaker pairing when going toward the drip-lines than standard DDDI depending on the total density and adjusted around the valley of stability. One can expect these properties of the force to significantly influence matter densities, pair densities, individual excitation spectra, low-energy vibrational collective modes, rotational properties and the odd-even mass differences in exotic nuclei as well as the position of the neutron drip-line. It was shown that the pairing gap is extremely sensitive to the details of the force when dealing with neutron rich nuclei~\cite{doba2,bennaceur}. Also, while the position of the proton drip-line appears to be quite robust, the position of the neutron drip-line was shown to be shifted by up to 25 mass units depending on whether the so-called volume or surface pairing force was used~\cite{bennaceur}. This is one of the achievements of the present work to propose a pairing interaction whose ab-initio derivation should make its extrapolated use to exotic systems reliable and subject to less uncertainty.

At the mean-field level, this ab-initio character deals with the bare force. This is the first essential piece since, unlike for normal superconductors in condensed matter, the bare NN force provides pairing between the constituents of the nucleus. However, the question arises of its contribution compared to the pairing generated by collective effects. Recently, the BCS gap provided by the realistic $AV18$ $NN$ force was shown to account for only half of the experimental odd-even mass staggering in $^{120}Sn$~\cite{barranco}. This was interpreted as a necessity to go beyond the mean-field to introduce the off-shell nucleon propagation associated with the particle-vibration coupling and the induced interaction generated by the exchange of the same surface vibrations between time-reversed states. Including those processes, the experimental odd-even mass staggering was reproduced in $^{120}Sn$~\cite{barranco}. This exchange of surface vibrations should provide the effective pairing force with an additional surface-peaked character and some isospin dependence due to the appearance of new vibrational modes toward the neutron drip-line. On the other hand, the same category of diagrams is known to decrease the gap in the bulk~\cite{dean,clark2,baldo,shen,schulze2}; even if the corresponding dressing of the vertex could change significantly when going from neutron matter to symmetric matter~\cite{heiselberg4}. The latter would add up another dependence of the interaction on the nuclear asymmetry. Thus, the situation concerning the net influence of beyond-mean-field effects on pairing in finite nuclei is unclear. Let us repeat that, as already discussed in section~{analysis}, calculations performed with the Gogny interaction enforce the idea that medium renormalizations on top of the bare force in the $T=1$ channel should not be that large when going to finite systems. 

However, one has to consider to more elements before concluding. First, the effect of three-body force on pairing should be treated if one works with microscopic interactions as proposed here. The three-body force has been shown to decrease the gap in infinite matter non-negligibly for $k_{F} \geq 0.8 \, fm^{-1}$~\cite{zuo}. Including the three-body force in the pairing channel could also be necessary to reproduce delicate phenomena such as the odd-even staggering and the kinks of differential charge radii~\cite{fayans2}. Second, one has to include the Coulomb force in the proton-proton pairing channel. From the comparison between proton-proton and neutron-neutron scattering phase-shifts in the $^{1}S_0$ channel~\cite{wiringa2}, one would not expect a strong anti-pairing effect from the Coulomb interaction. This is particularly true since the very low energy regime where the Coulomb force is active does not seem to matter so much when dealing with pairing as discussed in section~\ref{results}. However, it was shown that the self-consistency of the HFB calculations makes the Coulomb anti-pairing effect quite significant in finite nuclei~\cite{anguiano2}.

Thus, pairing correlations and details of the effective pairing interaction are far from being understood in nuclei. Systematic microscopic calculations including all previously mentioned effects are required. However, this is unconceivable in finite nuclei at this stage. As a first step, performing systematic HFB calculations with our effective interaction offers a unique opportunity to understand in detail the contribution of the bare force to pairing in nuclei.

Let us now discuss the range of the effective pairing force. Indeed, it would be nice to have an interaction allowing for large scale microscopic calculations of nuclear masses~\cite{goriely1,goriely2}. Performing self-consistent mean-field or beyond-mean-field calculations, the size of the single-particle basis necessary to get converged results should be as small as possible to make such large scale calculations tractable. The required size is directly related to the range of the force in the pairing channel. For instance the shorter range of the Gogny force ($0.7 \, fm$) makes the convergence of the quantities related to pairing quite slow. One needs to include states up to $100 \, MeV$ in the quasi-particle spectrum to do so~\cite{grasso2}. It can be seen from Fig.~\ref{comparisonvlowk} that our effective interaction is softer than the Gogny force. The range to be compared with the $0.7 \, fm$ of Gogny is in the present case $\sqrt2 \alpha \approx 1 fm$. This translates into a kinetic energy of about $70 \, MeV$ which, once the depth of the single-particle potential has been subtracted ($\approx \, 50 \, MeV$), gives a value of the order of $20 \, MeV$. Taking the effect of the cut-off $(\rho_{j} + \rho_{l})$ in the gap equation into account, it should be sufficient to include single-particle states up to $10-20 \, MeV$ of positive energy in the canonical basis to get converged HFB calculations. Thus, this pairing interaction should be simple and soft enough to perform large scale calculations of nuclear masses, avoiding at the same time the problem related to the phenomenological choice of the cut-off dealing with the zero range of the DDDI~\cite{goriely2}. Checking this statement is the aim of a forthcoming publication~\cite{duguet7}. Note that the induced interaction is expected to be long ranged~\cite{schulze2,barranco}.

Once the softness of the pairing interaction has been established, it will be important to understand the significance of the range and the non-locality of the force and whether it needs to be treated explicitly in nuclei. It amounts to studying the influence of the non-locality of the pairing field. Being able to use the microscopic non-local finite-range interaction and a gap-equivalent local zero-range force within a single scheme offers a unique opportunity to answer such a question~\cite{duguet7}.

\section{Conclusions}
\label{Conclusions}

In this paper, we have proposed a microscopic effective interaction to describe pairing in the $^{1}S_0$ channel. The new features of the interaction are numerous. 

It possesses a clear link to the bare force.  The effective interaction provides the same pairing properties as the bare force at the mean-field level, as required from many-body theories. The gap at the Fermi energy obtained in infinite matter from the realistic AV18 interaction is perfectly reproduced by the new force. Going into further detail, the momentum dependence of the gap is also very well described in the density range of interest. 

The effective force is finite ranged, non local, total-momentum dependent and density dependent. While the effective interaction is almost constant for densities ranging from saturation to typical surface densities, it is strongly enhanced at very low density. The isoscalar and isovector density dependences of the pairing force are also obtained through the ab-initio derivation. While phenomenological density-dependent pairing interactions used so far depend on the total density, the one derived here depends on the density of the interacting nucleons only (i.e. protons {\it or} neutrons since the present work deals with like-particle pairing only). 

This effective pairing force is defined by recasting the gap equation written in terms of the bare force into a fully equivalent pairing problem. Through this rewriting procedure, the matrix elements of the effective force are provided with a natural cut-off $2 \, \rho_{m}$, where $\rho_{m}$ is a BCS occupation number. This makes the definition of zero-range approximations meaningful and no ad-hoc cut-off has to be additionally chosen and optimized. Performing such a zero-range approximation and asking for identical pairing gaps at the Fermi surface in infinite matter, the appropriate density-dependence of the zero-range force is obtained. This procedure, free of any phenomenological cut-off allows us to disentangle the roles played by the range and the density dependence of the pairing interaction. Surprisingly, the enhancement of the force at typical surface densities as compared to the saturation density is much more pronounced than for the finite range vertex. This result shows unambiguously that the surface character of usual zero-range forces is, to a large extent, a way of taking care of the range of the interaction. Precisely, the zero-range vertex is predicted to have a behavior between surface and volume. It also undergoes a large increase of its intensity at very low density.

The last essential feature of the force can be identified when going from infinite matter to finite nuclei. Indeed, dealing with a finite-range and non-local interaction is far from being trivial when solving the HFB equations in coordinate space. However, the particular analytical structure of our effective force makes such calculations possible within the so-called two-basis method. In fact, the corresponding computational cost is of the same order as for zero-range forces. Performing exploratory calculations in finite nuclei with this new interaction is the aim of a forthcoming publication~\cite{duguet7}.

\section{Acknowledgments}
\label{secremer}

This research was supported by the US Department of Energy (DOE), Nuclear Physics Division, under contract no. W-31-109-ENG-38. The author thanks the Department of Energy's Institute for Nuclear Theory at the University of Washington and the Service de Physique Th\'eorique at the CEA/Saclay for their hospitality and partial support during the elaboration of this work. The author is grateful to M. Baldo, G.-F. Bertsch, P. Bonche, S. Pieper, B. Sabbey and R. B. Wiringa for inspiring and useful discussions. The author is grateful to S. K. Bogner, P. Schuck and R. B. Wiringa for providing him with results of their calculations and to N. J. Hammond for the proof reading of the manuscript.

\clearpage

%Bibliographie


\begin{thebibliography}{99}


\bibitem% [doba3]
          {doba3}
J. Dobaczewski and W. Nazarewicz, Prog. Theor. Phys. Suppl. {\bf 146} (2003) 70

\bibitem% [dean] 
          {dean}
D. J. Dean, M. Hjorth-Jensen, Rev. Mod. Phys. {\bf 75} (2003) 607 and references therein

\bibitem% [sauls] 
          {sauls}
J. A. Sauls, {\it Timing Neutron Stars}, ed. by H. Ogelman and E. P. J. van den Heuvel, (Dordrecht, Kluwer, 1989), 457 

\bibitem% [heiselberg3] 
          {heiselberg3}
H. Heiselberg and M. Hjorth-Jensen, Phys. Rep. {\bf 328} (2000) 237

\bibitem% [ring1] 
          {ring1}
P. Ring and P. Schuck, {\it The Nuclear Many-Body Problem} (Springer-Verlag, New-York, 1980)

\bibitem% [michael] 
          {michael}
M. Bender, P.-H. Heenen and P.-G. Reinhard, Rev. Mod. Phys. {\bf 75} (2003) 121

\bibitem% [hjorthjensen] 
          {hjorthjensen}
M. Hjorth-Jensen, T. T. S. Kuo and E. Osnes, Phys. Rep. {\bf 261} (1995) 125

\bibitem% [gorkov]
	  {gorkov}
L. P. Gorkov, Sov. Phys. JETP {\bf 34} (1958) 505

\bibitem% [bogo] 
          {bogo}
N. N. Bogolyubov, Soviet Phys. JETP {\bf 7} (1958) 41; N. N. Bogolyubov, Soviet Phys. Usp. {\bf 2} (1959) 236; N. N. Bogolyubov and V. G. Soloviev, Soviet Phys. Doklady {\bf 4} (1959) 143

\bibitem% [mehta1] 
          {mehta1}
M. L. Mehta, Th\`ese, Paris (1961), unpublished

\bibitem% [henley] 
          {henley}
E. M. Henley and L. Wilets, Phys. Rev. {\bf 133} (1964) B1118

\bibitem% [baldo3]
          {baldo3}
M. Baldo, J. Cugnon, A. Lejeune and U. Lombardo, Nucl Phys. {\bf A515} (1990) 409

\bibitem% [duguet9]
	  {duguet9}
T. Duguet, Phys. Rev. {\bf C67} (2003) 044311

\bibitem% [kohn1] 
          {kohn1}
P. Hohenberg and W. Kohn, Phys. Rev. {\bf 136} (1964) 864

\bibitem% [kohn2] 
          {kohn2}
W. Kohn and L. J. Sham, Phys. Rev. {\bf 140} (1965) 1133

\bibitem% [kucharek1]
          {kucharek1}
H. Kucharek and P. Ring, Z. Phys. {\bf A339} (1991) 23

\bibitem% [brett1] 
          {brett1}
F. B. Guimaraes, B. V. Carlson and T. Frederico, Phys. Rev. {\bf C54} (1996) 2385

\bibitem% [brett2] 
          {brett2}
B. V. Carlson, T. Frederico and F. B. Guimaraes, Phys. Rev. {\bf C56} (1997) 3097

\bibitem% [chen]
          {chen}
J. Chen, P. Zhuang and J. Li, preprint nucl-th/0309033

\bibitem% [dech]
	  {dech}
J. Decharg\'e and D. Gogny, Phys. Rev. {\bf C21} (1980) 1568

\bibitem% [bertsch]
          {bertsch}
G. F. Bertsch and H. Esbensen, Ann. Phys. (NY) {\bf 209} (1991) 327

\bibitem% [rigol]
	  {rigol}
C. Rigollet, P. Bonche, H. Flocard and P.-H. Heenen, Phys. Rev. {\bf C59} (1999) 3120

\bibitem% [duguet2]
	  {duguet2}
T. Duguet, P. Bonche and P.-H. Heenen, Nucl. Phys. {\bf A679} (2001) 427

\bibitem% [duguet6] 
          {duguet6}
T. Duguet, P. Bonche, P.-H. Heenen and  J. Meyer, Phys. Rev. {\bf C65} (2002) 014310

\bibitem% [duguet61] 
          {duguet61}
T. Duguet, P. Bonche, P.-H. Heenen and  J. Meyer, Phys. Rev. {\bf C65} (2002) 014311

\bibitem% [terasaki2]
	  {terasaki2}
J. Terasaki, F. Barranco, P.-F. Bortignon, R. A. Broglia and E. Vigezzi, nucl-th/0109056

\bibitem% [michael1]
	  {michael1}
M. Bender, P. Bonche, T. Duguet and P.-H. Heenen, Nucl. Phys. {\bf A723} (2003) 354

\bibitem% [tajima2]
	  {tajima2}
N. Tajima, P. Bonche, H. Flocard, P.-H. Heenen and M. S. Weiss, Nucl. Phys. {\bf A551} (1993) 434

\bibitem% [fayans2] 
          {fayans2}
D. Zawischa, Phys. Lett. {\bf B155} (1985) 309; D. Zawischa, U. Regge and R. Stapel, Phys. Lett. {\bf B185} (1987) 299; U. Regge and D. Zawischa, Phys. Rev. Lett. {\bf 61} (1988) 149; S. Fayans, S. V. Tolokonnikov, E. L. Trykov and D. Zawischa, Phys. Lett. {\bf B338} (1994) 1; S. Fayans and D. Zawischa, Phys. Lett. {\bf B383} (1996) 19; S. Fayans, S. V. Tolokonnikov, E. L. Trykov and D. Zawischa, Nucl. Phys. {\bf A676} (2000) 49

\bibitem% [doba4]
          {doba4}
J. Dobaczewski, W. Nazarewicz and P. G. Reinhard, Nucl. Phys. {\bf A693} (2001) 361

\bibitem% [doba2]
          {doba2}
J. Dobaczewski, W. Nazarewicz and M. V. Stoitsov, Eur. Phys. J. {\bf A15} (2002) 21

\bibitem% [doba7]
          {doba7}
J. Dobaczewski, W. Nazarewicz and M. V. Stoitsov, Proceedings of the NATO Advanced Research Workshop {\it The Nuclear Many-Body Problem 2001}, Brijuni, Croatie, 2-5 Juin, 2001; nucl-th/010973

\bibitem% [fallon]
	  {fallon} 
P. Fallon, P.-H. Heenen, W. Satula, R. M. Clark, F. S. Stephens, M. A. Deleplanque, R. M. Diamond, I. Y. Lee, A. O. Macchiavelli and K. Vetter, Phys. Rev. C60 (1999), 044301

\bibitem% [ph]
	  {ph}
P.-H. Heenen, J. Terasaki, P. Bonche, H. Flocard, and J. Skalski, Czech. J. Phys. {\bf 48}, 671 (1998)

\bibitem% [goriely1]
	  {goriely1}
S. Goriely, F. Tondeur and J. M. Pearson, At. Data Nucl. Data Tables {\bf 77} (2001) 311

\bibitem% [samyn]
	  {samyn}
S. Samyn, S. Goriely, P.-H. Heenen, J. M. Pearson and F. Tondeur, Nucl. Phys. {\bf A700} (2002) 142

\bibitem% [goriely2]
	  {goriely2}
S. Goriely, S. Samyn, P.-H. Heenen, J. M. Pearson and F. Tondeur, Phys. Rev. {\bf C66} (2002) 024326

\bibitem% [garrido1]
          {garrido1}
E. Garrido, P. Sarriguren, E. Moya de Guerra, U. Lombardo, P. Schuck and H. J. Schulze, Phys. Rev. {\bf C63} (2001) 037304

\bibitem% [garrido2]
          {garrido2}
E. Garrido, P. Sarriguren, E. Moya de Guerra and P. Schuck, Phys. Rev. {\bf C60} (1999) 064312

\bibitem% [migdal]
	  {migdal}
A. B. Migdal, {\it Theory of Finite Fermi Systems and Applications to Atomic Nuclei} (Interscience, New-York, 1967)

\bibitem% [vauth] 
          {vauth}
D. Vautherin and D. M. Brink, Phys. Rev. {\bf C5} (1972) 626

\bibitem% [doba8]
	  {doba8}
J. Dobaczewski, W. Nazarewicz, T. R. Werner, J.-F. Berger, C. R. Chinn and J. Decharg\'e, Phys. Rev. {\bf C53} (1996) 2809

\bibitem% [duguet8]
	  {duguet8}
T. Duguet and P. Bonche, Phys. Rev. {\bf C67} (2003) 054308

\bibitem% [bulgac1]
	  {bulgac1}
A. Bulgac, Phys. Rev. {\bf C65} (2002) 051305

\bibitem% [bulgac2]
	  {bulgac2}
Y. Yu and A. Bulgac, Phys. Rev. Lett. {\bf 90} (2003) 222501

\bibitem% [doba6]
	  {doba6}
J. Dobaczewski, H. Flocard and J. Treiner, Nucl. Phys. {\bf A422} (1984) 103

\bibitem% [gall]
	  {gall}
B. Gall, P. Bonche, J. Dobaczewski, H. Flocard and P.-H. Heenen, Z. Phys. {\bf A348} (1994) 183

\bibitem% [terasaki3]
	  {terasaki3}
J. Terasaki, P.-H. Heenen, H. Flocard and P. Bonche, Nucl. Phys {\bf A600} (1996) 371

\bibitem% [clark2]
	  {clark2}
J. W. Clark, C.-G. K\"{a}llman, C.-H. Yang and D. A. Chakkalakal, Phys. Lett. {\bf B61} (1976) 331

\bibitem% [baldo]
	  {baldo}
M. Baldo, U Lombardo, H.-J. Schulze and Z. Wei, Phys. Rev. {\bf C66} (2002) 054304

\bibitem% [shen] 
          {shen}
C. Shen, U. Lombardo and P. Schuck, preprint nucl-th/0212027

\bibitem% [schulze2] 
          {schulze2}
H.-J. Schulze, J. Cugnon, A. Lejeune, M. Baldo and U. Lombardo, Phys. Lett. {\bf B375} (1996) 1

\bibitem% [heiselberg4] 
          {heiselberg4}
H. Heiselberg, C. J. Pethick, H. Smith and L. Viverit, Phys. Rev. Lett. {\bf 85} (2000) 2418

\bibitem% [terasaki1]
	  {terasaki1}
J. Terasaki, F. Barranco, R. A. Broglia, E. Vigezzi and P. F. Bortignon, Nucl. Phys. {\bf A697} (2002) 127

\bibitem% [barranco]
	  {barranco}
F. Barranco, R. A. Broglia, G. Col\`o, E. Vigezzi and P.-F. Bortignon, nucl-th/0304049

\bibitem% [elgar1] 
          {elgar1}
O. Elgaroy and M. Hjorth-Jensen, Phys. Rev. {\bf C57} (1998) 1174

\bibitem% [teramond] 
          {teramond}
G. F. de T\'eramond and B. Gabioud, Phys. Rev. {\bf C36} (1987) 691

\bibitem% [gonzales] 
          {gonzales}
D. E. Gonz\'alez Trotter, F. Salinas, Q. Chen, A. S. Crowell, W. Gl\"{o}ckle, C. R. Howell, C. D. Roper, D. Schmidt, I. Slaus, H. Tang, W. Tornow, R. L. Walter, H. Witala and Z. Zhou, Phys. Rev. Lett. {\bf 83} (1999) 3788

\bibitem% [koester] 
          {koester}
L. Koester and W. Nistler, Z. Phys. {\bf A272} (1975) 189 

\bibitem% [bergervoet] 
          {bergervoet}
J. R. Bergervoet, P. C. van Campen, W. A. van der Sanden and J. J. de Swart, Phys. Rev. {\bf C38} (1988) 15

\bibitem% [brown] 
          {brown}
G. E. Brown and A. D. Jackson, {\it The Nucleon-Nucleon Interaction} (North-Holland, Amsterdam, 1976)

\bibitem% [emery2] 
          {emery2}
V. J. Emery and A. M. Sessler, Phys. Rev. {\bf 119} (1960) 248

\bibitem% [kwong]
	  {kwong}
N. H. Kwong and H. S. K\"{o}hler, Phys. Rev. {\bf C55} (1997) 1650

\bibitem% [kennedy]
	  {kennedy}
R. Kennedy, L. Wilets and E. M. Henley, Phys. Rev. {\bf 133} (1964) B1131

\bibitem% [bardeen]
	  {bardeen}
J. Bardeen, L. N. Cooper and J. R. Schrieffer, Phys. Rev. {\bf 106} (1957) 162; Phys. Rev. {\bf 108} (1957) 1175

\bibitem% [galitskii] 
          {galitskii}
V. M. Galitskii, Sov. Phys. JETP {\bf 34} (1958) 104

\bibitem% [larkin] 
          {larkin}
A. I. Larkin and Yu. N Ovchinnikov, Zh. Eksp. Teor. Fiz. {\bf 47} (1964) 1136 [Sov. Phys. JETP {\bf 20} (1965) 762]

\bibitem% [fulde] 
          {fulde}
P. Fulde and R. A. Ferrell, Phys. Rev. {\bf 135} (1964) 550

\bibitem% [sedrakian] 
          {sedrakian}
A. Sedrakian, Phys. Rev. {\bf C63} (2001) 025801

\bibitem% [bozek] 
          {bozek}
P. Bozek, Nucl. Phys {\bf A657} (1999) 187;  Phys. Rev. {\bf C65} (2002) 034327

\bibitem% [brueck] 
          {brueck}
K. A. Brueckner, C. A. Levinson and H. M. Mahmoud, Phys. Rev. {\bf 95} (1954) 217; K. A. Brueckner and C. A. Levinson, Phys. Rev. {\bf 97} (1954) 1344; K. A. Brueckner, J. L. Gammel and H. Weitzner, Phys. Rev. {\bf 110} (1958) 431;  K. A. Brueckner and D. T. Goldman, Phys. Rev. {\bf 117} (1960) 207; K. A. Brueckner, A. M. Lockett and M. Rotenberg, Phys. Rev. {\bf 121} (1961) 255

\bibitem% [skyrme]
	  {skyrme}
T. H. R. Skyrme, Philos. Mag. {\bf 1} (1956) 1043; T. H. R. Skyrme, Nucl. Phys. {\bf 9} (1959) 615

\bibitem% [wiringa2] 
          {wiringa2}
R. B. Wiringa, V. G. J. Stocks and R. Schiavilla, Phys. Rev. {\bf C51} (1995) 38

\bibitem% [khodel] 
          {khodel}
V. A. Khodel, V. V. Khodel and J. W. Clarck, Nucl. Phys. {\bf A598} (1996) 390

\bibitem% [lombardo1] 
          {lombardo1}
U. Lombardo and H.-J. Schulze, Lect. Notes Phys. {\bf 578} (2001) 30

\bibitem% [bogner]
          {bogner}
S. Bogner, T. T. S. Kuo and L. Coraggio, Nucl. Phys. {\bf A684} (2001) 432c

\bibitem% [steele1]
          {steele1}
J. V. Steele and R. J. Furnstahl, Nucl. Phys. {\bf A663} (2000) 999

\bibitem% [steele2]
          {steele2}
J. V. Steele, preprint nucl-th/0010066

\bibitem% [bogner2]
          {bogner2}
S. Bogner, T. T. S. Kuo and A. Schwenk, preprint nucl-th/0305035

\bibitem% [sedrakian1]
          {sedrakian1}
A. Sedrakian, T. T. S. Kuo, H. M\"{u}ther and P. Schuck, preprint nucl-th/0308068

\bibitem% [d1s]
          {d1s}
J.-F. Berger, M. Girod and D. Gogny, Comp. Phys. Comm. {\bf 63} (1991) 365

\bibitem% [anguiano]
          {anguiano}
M. Anguiano, J.L. Egido and L.M. Robledo, Nucl.Phys. {\bf A696} (2001) 467

\bibitem% [peru]
          {peru}
S. P\'eru, M. Girod and J.-F. Berger, Eur. Phys. J. {\bf A9} (2000) 35

\bibitem% [egido1]
          {egido1}
J. L. Egido and L. M. Robledo, Phys. Rev. Lett. {\bf 70} (1993) 2876

\bibitem% [girod]
          {girod}
M. Girod and B. Grammaticos, Phys. Rev. {\bf C27} (1983) 2317

\bibitem% [kucharek2]
          {kucharek2}
H. Kucharek, P. Ring and P. Schuck, Z. Phys. {\bf A334} (1989) 119

\bibitem% [wiringa1] 
          {wiringa1}
R. B. Wiringa, private communication

\bibitem% [bozek2] 
          {bozek2}
P. Bozek, Phys. Rev. {\bf C62} (2000) 054316

\bibitem% [bloch] 
          {bloch}
C. Bloch and A. Messiah, Nucl. Phys. {\bf 39} (1962) 95

\bibitem% [bozek1]
          {bozek1}
P. Bozek, Phys. Lett. {\bf B551} (2003) 93

\bibitem% [esbensen] 
          {esbensen}
H. Esbensen, G. F. Bertsch and K. Hencken, Phys. Rev. {\bf C56} (1997) 3054

\bibitem% [marini] 
          {marini}
M. Marini, F. Pistolesi aand G. C. Strinati, Eur. Phys. J. {\bf B1} (1998) 151

\bibitem% [papenbrock] 
          {papenbrock}
T. Papenbrock and G. F. Bertsch, Phys. Rev. {\bf C59} (1999) 2052

\bibitem% [thouless]
	  {thouless}
D. J. Thouless, Ann. Phys. (N.Y.) {\bf 10} (1960) 553

\bibitem% [bishop1] 
          {bishop1}
R. F. Bishop, M. R. Strayer and J. M. Irvine, Phys. Rev. {\bf A10} (1974) 2423

\bibitem% [dickhoff]
	  {dickhoff}
W. H. Dickhoff, C. C. Gearhart, E. P. Roth, A. Polls, and A. Ramos, Phys. Rev. {\bf C60} (1999) 064319

\bibitem% [balian] 
          {balian}
R. Balian and M. L. Metha, Nucl. Phys. {\bf 31} (1962) 587

\bibitem% [elgar2] 
          {elgar2}
O. Elgaroy, L. Engvik, M. Hjorth-Jensen and E. Osnes, Nucl. Phys. {\bf A604} (1996) 466

\bibitem% [baldo0]
          {baldo0}
M. Baldo, U. Lombardo, E. E. Saperstein and M. V. Zverev, Nucl. Phys. {\bf A628} (1998) 503

\bibitem% [baldo1]
          {baldo1}
M. Baldo, U. Lombardo, E. E. Saperstein and M. V. Zverev, Phys. Lett. {\bf B477} (2000) 410

\bibitem% [doba1]
	  {doba1}
J. Dobaczewski, Plenary talk at the INPC'01, Berkeley, USA, 30 July - 3 August, 2001

\bibitem% [bonche1] 
          {bonche1}
P. Bonche, H. Flocard, P.-H. Heenen, S. J. Krieger and M. S. Weiss, Nucl. Phys. {\bf A443} (1985) 39

\bibitem% [chab] 
          {chab}
E. Chabanat, P. Bonche, P. Haensel, J. Meyer and R. Schaeffer, Nucl. Phys. {\bf A627} (1997) 710; Nucl. Phys. {\bf A635} (1998) 231; Nucl. Phys. {\bf A643} (1998) 441(E)

\bibitem% [schulze] 
          {schulze}
H.-J. Schulze, A. Schnell, G. R\"{o}pke and U. Lombardo, Phys. Rev. {\bf C55} (1997) 3006

\bibitem% [baldo4]
          {baldo4}
M. Baldo, O. Elgaroy, L. Engvik, M. Hjorth-Jensen and H.-J. Schulze, Phys. Rev. {\bf C58} (1998) 1921

\bibitem% [lombardo2] 
          {lombardo2}
U. Lombardo, H.-J. Schulze and W. Zuo, Phys. Rev. {\bf C59} (1999) 2927

\bibitem% [zuo]
          {zuo}
W. Zuo, U. Lombardo, H.-J. Schulze and C. W. Shen, Phys. Rev. {\bf C66} (2002) 037303

\bibitem% [delion]
	  {delion}
D. S. Delion, M. Baldo and U. Lombardo, Nucl. Phys {\bf A593} (1995) 151

\bibitem% [kucharek]
          {kucharek}
H. Kucharek, P. Ring, P. Schuck, R. Bengtsson and M. Girod, Phys. Lett. {\bf B216} (1989) 249

\bibitem% [bulgac3]
	  {bulgac3}
A. Bulgac and Y. Yu,  10$^{th}$ Nuclear Physics Workshop Marie and Pierre Curie, Kazimierz Dolny, 24-28 September 2003. To appear in a special issue of Int. Journ. Mod. Phys. E; preprint nucl-th/0310066

\bibitem% [grasso2]
	  {grasso2}
M. Grasso, N. Van Giai and N. Sandulescu, Phys. Lett. {\bf B535} (2002) 103

\bibitem% [terasaki4]
	  {terasaki4}
J. Terasaki, P.-H. Heenen, P. Bonche, J. Dobaczewski and H. Flocard, Nucl. Phys {\bf A593} (1995) 1

\bibitem% [yamagami]
	  {yamagami}
M. Yamagami, K. Matsuyanagi and M. Matsuo, Nucl. Phys {\bf A693} (2001) 579

\bibitem% [teran]
	  {teran}
E. T\'eran, V. E. Oberacker and A. S. Umar, Phys. Rev. {\bf C67} (2003) 064314

\bibitem% [bennaceur]
	  {bennaceur}
K. Bennaceur, P. Bonche and J. Meyer, C. R. Physique {\bf 4} (2003) 555

\bibitem% [anguiano2]
          {anguiano2}
M. Anguiano, J.L. Egido and L.M. Robledo, Nucl.Phys. {\bf A683} (2001) 227

\bibitem% [duguet7]
	  {duguet7}
T. Duguet,  G.-F. Bertsch, B. Sabbey and P. Bonche, in preparation


\end{thebibliography}
\end{document}